\documentstyle[12pt,epsf]{article}
 
\begin{document}

\thispagestyle{empty}
\renewcommand{\thefootnote}{\fnsymbol{footnote}}
\setcounter{page}{0}
\begin{flushright}
CERN-TH/96-362\\
NORDITA 96/79-P\\
January 1997\\
hep-ph/9701309
\end{flushright}
\vspace*{1cm}
\centerline{\Large\bf Phenomenology of Power Corrections in}
\vspace*{0.1cm}
\centerline{\Large\bf Fragmentation Processes in $e^+ e^-$ Annihilation}
\vspace*{1.3cm}
\centerline{{\sc M.~Beneke}}
\bigskip
\centerline{\sl Theory Division, CERN,}
\centerline{\sl CH-1211 Geneva 23, Switzerland}
\vskip0.6truecm
\centerline{{\sc V.M.~Braun} and {\sc L.~Magnea}
\footnote{On leave of absence from Universit\`{a} di Torino, Torino, Italy.}
}
\bigskip
\centerline{\sl NORDITA, Blegdamsvej 17, DK--2100 Copenhagen, Denmark}
 
\vspace*{0.6cm}
\centerline{\bf Abstract}
\vspace*{0.2cm}
\noindent 
We analyse power corrections to longitudinal and transverse fragmentation 
processes in $e^+ e^-$ annihilation, based on the assumption of 
ultraviolet dominance of power corrections. Under this assumption, 
we determine the dependence of power corrections on the scaling 
variable $x$ from the infrared renormalon asymptotics of leading power 
coefficient functions. Our results suggest that the longitudinal and 
transverse gluon fragmentation coefficient functions receive corrections 
of order $1/(x Q)^2$. The power expansion breaks down at $x<\Lambda/Q$ and 
has to be resummed. This resummation leads to $1/Q$ corrections to the 
longitudinal and transverse cross section, which cancel for the total cross 
section. We provide a simple parametrization of the $x$ dependence of 
$1/Q^2$ corrections to fragmentation processes and investigate perturbative 
corrections to the longitudinal cross section in higher orders, in view of 
a determination of the strong coupling.

\noindent 
\renewcommand{\thefootnote}{\arabic{footnote}}
\setcounter{footnote}{0}

\newpage
\section{Introduction}

Inclusive single-particle production in 
$e^+ e^-$ collisions, $e^+ e^-\to\gamma^*,Z^0\to H(p)+X$, tests 
scaling violations in the time-like region and can be used to 
measure the strong coupling $\alpha_s$ \cite{DELPHI,OPAL,ALEPH}. Additional 
insight can be gained from a measurement of the angular dependence 
of the detected hadron \cite{OPAL,ALEPH}, since, for example, gluon 
fragmentation enters the longitudinal cross section as a leading 
contribution. The differential cross 
section can be expressed as \cite{NAS94}
\begin{eqnarray}
\label{def}
\frac{d^2\sigma^H}{dx d\cos\theta}(e^+e^-\to HX)\!&=&\! 
\frac{3}{8}\,(1+\cos^2\theta)\,\frac{d\sigma_T^H}{dx}(x,Q^2)
+\frac{3}{4}\sin^2\theta\, \frac{d\sigma_L^H}{dx}(x,Q^2)
\nonumber\\
&&\hspace*{-3cm}+\,\frac{3}{4}\cos\theta\, \frac{d\sigma_A^H}{dx}(x,Q^2) .
\end{eqnarray}
We defined $x=2 p\cdot q/q^2$, where $p$ is the momentum of $H$ and $q$ 
the intermediate gauge boson momentum. $Q^2=q^2$ denotes the center-of-mass 
energy squared and $\theta$ the angle between the hadron and the beam axis. 
The angular decomposition in (\ref{def}) corresponds to 
the contributions from longitudinal and transverse polarizations of the 
intermediate gauge boson and from $\gamma^* - Z^0$ interference. In the 
following, we will not be concerned with the asymmetric contribution and 
with quark mass effects. Neglecting quark masses, $(1/\sigma_0)\, d\sigma^
H_{T/L}/dx$ (where $\sigma_0$ is the Born total annihilation cross section) 
is independent of electroweak couplings and the longitudinal cross 
section is suppressed by $\alpha_s$. The most precise measurements of 
fragmentation functions refer to a sum over all charged hadrons. In the 
following, we drop the superscript `$H$' when a sum over all hadrons $H$ 
is understood. The conversion from charged hadrons to all hadrons is 
considered as an `experimental problem'. 

It follows from the factorization properties of perturbative QCD that 
the `structure functions' in (\ref{def}) are convolutions of 
perturbative coefficient functions $C_P^i(x,Q^2/\mu^2)$ ($P=T,L,A$) 
and parton fragmentation functions $D_i^H(x,\mu)$ ($i=q,\bar{q},g$)
\footnote{For $i=q$ (or $\bar{q}$) we always imply that all light 
quark flavours are summed over. In the context of power corrections 
a quark is considered `light' if its mass is smaller than 
$\Lambda$. Charm and bottom quarks must be treated separately, 
even if the center-of-mass energy is much larger than their masses.}, 
\begin{equation}
\label{DGLAP}
\frac{d\sigma_P^H}{dx}(x,Q^2) =\sum_i\int_x^1\frac{dz}{z}\, 
C_P^i(z,Q^2/\mu^2)\,D_i^H(x/z,\mu) ,
\end{equation} 
up to corrections suppressed by some power of $\Lambda/Q$, where 
$\Lambda$ is the QCD scale parameter. The fragmentation functions 
have to be determined experimentally, but once measured, they can be 
used to predict fragmentation processes in other hard collisions. 
Both coefficient and fragmentation functions depend on the chosen 
factorization scheme and scale $\mu$. We choose dimensional 
regularization with $\overline{\mbox{MS}}$ subtractions. Then the 
coefficient functions are obtained as the partonic cross sections 
with poles minimally subtracted. The fragmentation functions satisfy 
time-like evolution equations with kernels known to next-to-leading 
order \cite{FUR80}. The coefficient functions in (\ref{DGLAP}) have 
been computed to order $\alpha_s^2$ \cite{RIJ96}. 

The evolution kernels and perturbative corrections to coefficient 
functions cause scaling violations that are logarithmic in the 
center-of-mass energy. As in deep-inelastic scattering (DIS), 
multi-parton correlations, not taken into account in (\ref{DGLAP}),  
lead to scaling violations that scale as some power of $\Lambda/Q$. 
Such corrections can nevertheless be important, as they can be enhanced 
in specific kinematic regions. Such a situation is well known for DIS, 
where for $x_{Bj}\to 1$ all partons must be highly correlated in order 
that all momentum can be transferred to a single parton. In this 
region, the twist expansion breaks down (see for example Ref.~\cite{BBL}). 
Unlike DIS, power corrections (in analogy with DIS we 
use `higher-twist correction' synonymously) in fragmentation 
processes have rarely been studied theoretically \cite{BB91}. 
The main difference comes from the applicability of the operator 
product expansion to DIS, which allows one to express the moments of 
multi-parton correlation functions \cite{HT} in terms of matrix elements of 
local operators. Such a classification seems to be more difficult 
for fragmentation, so that even the power behaviour of higher-twist 
corrections is not well established. Although the collinear (light-cone) 
expansion for fragmentation functions \cite{BB91} is similar to that of 
structure functions in DIS and leads to the introduction of 
generalized multi-parton correlation functions for fragmentation that 
parametrize $1/Q^2$ corrections, the 
moments of these correlation functions remain non-local quantities 
and a separation of short and long distances is not straightforward. 
Phenomenological hadronization models are at variance with the 
light-cone expansion of Ref.~\cite{BB91}, as they typically lead to 
$1/Q$ power corrections. The question of $1/Q$ versus $1/Q^2$ 
power behaviour is experimentally unsettled \cite{ALEPH}. 

The second moment of single-particle inclusive cross sections summed 
over all hadrons is of particular interest. Because of the energy 
conservation sum rule 
\begin{equation}
\sum_H \int\limits_0^1\!dx\,x \,D^H_i(x,\mu)=1, 
\end{equation}
the fragmentation functions disappear from the integrals
\begin{equation}
\sigma_P \equiv  \sum_H\frac{1}{2}\int_0^1\! 
dx \, x\frac{d\sigma_P^H}{dx}
= \sum_i\frac{1}{2}\int_0^1\! dx \,x\,C_P^i(x),
\label{moment2}
\end{equation}
which can therefore be calculated in perturbation theory up to corrections 
suppressed by some power of $\Lambda/Q$. The normalization is such that 
the total cross section $\sigma_{tot}=\sigma_L+\sigma_T$. For 
the longitudinal cross section \cite{RIJ96}
\begin{equation}
\label{sigmal}
\sigma_L=\sigma_0\left[
\frac{\alpha_s}{\pi}+(14.583-1.028N_f)\left(\frac{\alpha_s}{\pi}\right)^2
\!+\ldots\right],
\end{equation}
where $\alpha_s\equiv\alpha_s(Q)$ and $N_f$ is the number of active fermion 
flavours. While dispersion relations relate the total cross section to the 
operator product expansion of a current correlation function, the power 
corrections to the longitudinal or transverse cross section cannot be 
inferred from such methods. From the theoretical point of view, the 
longitudinal and transverse cross sections can be considered as event 
shapes, and little has been known about their power corrections 
until recently \cite{MW}--\cite{revs}. 
In Ref.~\cite{WEB94} a $1/Q$ power correction to the longitudinal cross 
section was suggested as a consequence of phase-space reduction in the 
one-gluon emission diagram, when calculated with a massive gluon. 
At first sight, this conclusion seems to be 
again in conflict with the expectation 
\cite{BB91} that fragmentation functions need to be corrected only 
at order $1/Q^2$.

The present paper is devoted to a theoretical 
analysis of power corrections in fragmentation processes.\footnote{
A preliminary account appeared in Ref.~\cite{BBM}.} We collect 
results of phenomenological character, which we hope can provide 
useful guidance in analysing fragmentation data collected at 
$e^+ e^-$ colliders and in extracting the strong coupling from the 
longitudinal cross section as well as, possibly, other event shape 
observables. The interpretation of our results in operator language 
\cite{BB91} will be given in a subsequent paper.

In Sect.~2 we summarize the method, based on the infrared sensitivity 
of Feynman graphs, that allows us to estimate the $x$ dependence of 
power corrections and clarify the assumptions that enter this approach.
In Sect.~3, we give results for the power corrections to the quark 
and gluon coefficient functions. As will be explained, these power 
corrections factorize, so that the final estimate of power corrections 
to the fragmentation cross sections is given as a convolution with the 
leading twist fragmentation functions. 
A large part of Sect. 3 explores numerically the differences in various 
implementations of the method. Their comparison suggests an effective 
parametrization of $1/Q^2$ power corrections, presented in Sect. 3.5, 
which captures the gross features of the $x$ dependence. This
parametrization could be applied to LEP data \cite{DELPHI,OPAL,ALEPH}, 
as a substitute for phenomenological hadronization corrections obtained 
from Monte Carlo programs. We also find that the power expansion 
is singular in the region where the detected hadron is soft and emanates 
from a gluon jet. For small values of $x$, the effective expansion 
parameter is $(\Lambda/(x Q))^2$. The consequences of these small-$x$ 
singularities are pursued in Sect.~4, where we show that their 
resummation leads to a $1/Q$ power correction to the transverse and 
longitudinal cross section, thus resolving the apparent conflict with 
the light-cone expansion at fixed $x$. In Sect.~5 we argue that the 
large second-order correction in (\ref{sigmal}) is not accidental, but 
reflects the fact that the longitudinal cross section receives 
a $1/Q$ correction. This leads us to examine yet higher-order corrections 
in a certain approximation. We discuss their effect on a determination of 
$\alpha_s$ and the energy dependence of the longitudinal cross section 
with an eye on generic features that would equally apply to other 
event shape variables.

The results and conclusions of Sects. 3 and 4 overlap with those 
obtained independently by Dasgupta and Webber \cite{DAS96}, who 
use a somewhat different model for the gluon contribution to the 
fragmentation functions. We discuss this difference and other ambiguities 
inherent in the method in Sect.~3.

\section{Method and assumptions}

\subsection{Ultraviolet dominance}

Although the method we use to estimate the $x$ dependence of higher-twist 
corrections has previously been used \cite{DMW,STE96,DAS962} in DIS, we 
find it useful to repeat the main ideas and to spell out the assumptions. 
Since the method applies without conceptual difference to fragmentation 
and DIS, it might be helpful to have DIS in mind as an example, for 
which the language of higher-twist corrections is familiar and higher-twist 
corrections can be interpreted within the operator product expansion.

We start from the observation that the separation of leading-twist from 
higher-twist is not unique. 
This is not apparent in low-order perturbative 
calculations in the $\overline{\mbox{MS}}$ scheme. Imagine, however, that 
the factorization in transverse momenta that is implicit in (\ref{DGLAP}) 
were implemented by a rigid cut-off $\mu$, such that only contributions from 
transverse momenta $k_t>\mu$ were included in the coefficient 
function $C_P^i$. Then one would find a term $\ln Q^2/\mu^2$, whose 
cut-off dependence is cancelled by the $\mu$ dependence of leading-twist 
fragmentation functions $D_i^H$, and in addition power-like cut-off 
dependence, starting with $\mu^2/Q^2$, which is cancelled by higher-twist 
contributions. Therefore the leading-twist contribution as a whole 
depends on the prescription used to implement the cut-off, 
just as the separation of leading-twist coefficient and 
fragmentation functions in (\ref{DGLAP}) does. At first 
sight, such prescription dependence seems to be avoided in the 
$\overline{\mbox{MS}}$ scheme, because power-like dependence on the 
factorization scale $\mu$ does not exist. The problem reappears, however, 
because the  coefficient function now has a factorially divergent 
series expansion in $\alpha_s$ (referred to as infrared renormalon 
divergence). Summing the series again requires a prescription and the 
prescription-dependence is power-suppressed precisely as the cut-off 
dependence above. In both cases, the ultraviolet renormalization of 
higher-twist operators must be performed consistently with the definition 
of the leading-twist coefficient function. The sum of leading-twist and 
higher-twist contributions is then unique. 
For the case at hand, setting $\mu=Q$ in (\ref{DGLAP}), we can 
write the leading power ambiguity in the leading twist coefficient function
as 
\begin{equation}
\label{ambi}
\delta C_P^i(x)\propto A_{2,P}^i(x)\,\left(\frac{\Lambda}{Q}\right)^2
\end{equation}
times, possibly, logarithms of $\Lambda/Q$. The functions $A_{2,P}^i(x)$ are  
calculable in a certain approximation, as explained below. 

Whichever point of view one prefers, there are two immediate conclusions: 
first, from the phenomenological point of view, higher-order contributions 
in perturbation theory and higher-twist corrections are inseparable and 
should be described by one parameter; second, precisely for this 
reason, some information on higher-twist effects can be obtained from 
the infrared (or large-order) behaviour of perturbation theory.
If we could enumerate the higher-twist operators as in DIS, this piece 
of information would refer only to the ultraviolet regularization
properties of these higher-twist operators, with an $x$ dependence given 
by $A_{2,P}^i(x)$ in (\ref{ambi}).
In the following we will assume that the $x$ dependence of the entire 
higher-twist contribution is proportional to $A_{2,P}^i(x)$ and refer 
to this assumption as `ultraviolet dominance of higher-twist 
corrections'. Eq.~(\ref{DGLAP}) is then replaced by 
\begin{eqnarray}
\label{withpower}
\frac{d\sigma_P}{dx}(x,Q^2) &=&\sum_i\int_x^1\frac{dz}{z}\, 
\bigg[C_P^i(z,Q^2/\mu^2)+K_2^i A_{2,P}^i(z)\,\frac{\Lambda^2}{Q^2}
\nonumber\\
&&\,+K_4^i A_{4,P}^i(z)\,\frac{\Lambda^4}{Q^4}+\ldots\bigg]
\,D_i(x/z,\mu).
\end{eqnarray} 
The dots denote higher power corrections and we anticipated that 
only even powers in $1/Q$ occur. The functions $A_{2,P}^i(z)$ will be given 
in Sect.~3.5 and the $K_n^i$ are adjustable, $z$-independent 
constants that should be determined by comparison with experimental 
data. Note that here we slightly differ from previous formulations, 
where the constants are fixed in terms of an `effective coupling' 
\cite{DMW,DAS962} or equated to a universal constant \cite{STE96} 
(fixed by the relation to the IR renormalon ambiguity). 
In our approach, these constants can 
depend on the factorization scale and also the order of perturbation 
theory to which the leading-twist coefficient function $C_P^i$ has 
been calculated, since the added power corrections partly 
parametrize higher-order perturbative corrections as well. 
Naturally, the constants $K_n^i$ should be `of order 1'. We 
emphasize again that for $i=q$, a sum over light quark flavours 
($u$, $d$ and $s$) is already included in $A_{n,P}^i(z)$.

Arguments related to infrared cut-off behaviour or infrared renormalons 
have been used repeatedly \cite{revs} to determine the scaling of power 
corrections with $1/Q$. In this case, no new information is obtained 
for DIS, where the power behaviour is known from the operator product 
expansion (OPE). The dependence of power corrections on $x_{Bj}$ is not 
constrained by the OPE and is given in terms of multi-parton correlation 
functions, which are already too complex to be extracted from 
experiment. The ultraviolet dominance hypothesis provides tremendous 
simplifications, as the unknowns are reduced to a few constants rather 
than functions. Results that followed from applying this hypothesis to 
DIS structure functions have been found to reproduce experimental 
results on the $x_{Bj}$-dependence of higher-twist contributions 
unexpectedly well \cite{DMW,STE96,DAS962}. This empirical success 
encourages us to try the same idea for fragmentation processes.

The assumption that higher-twist corrections are proportional to their 
cut-off dependence (or renormalon ambiguity) does not become correct 
in any limit and therefore the resulting $x$ dependence should be 
considered as a model. One should not insist on this proportionality 
too literally. Rather, the idea is that if the ultraviolet contribution 
to the higher-twist correction varies rapidly with $x$, one may expect 
that the full higher-twist correction also does, up to some smooth 
function. This expectation is best motivated by an analogy. Suppose 
we calculated some quantity to order $\alpha_s^n$ in perturbation theory. 
It is customary to vary the scale $\mu$ to get an idea about the size of the 
next term in the expansion. Of course, the $\mu$-dependence gets exactly 
compensated, but if it is large (or small) the $\mu$-independent terms 
are also expected to be large (or small). In the same way, although 
renormalon ambiguities are ultimately cancelled and unphysical, their 
$x$ dependence should be indicative of the 
$x$ dependence of the full higher-twist 
contribution.

Related to the choice of scale is the fact that in certain kinematical 
regions the scale of the hard process turns out to be of the form 
$\varepsilon(x) Q$, with $\varepsilon(x)$ typically given by a power of
$x$ or of $1 - x$, because of phase-space restrictions for gluon emission.
The physical scale is thus parametrically smaller than $Q$,
and the power expansion is naturally organized in terms of 
$\Lambda/(\varepsilon(x) Q)$ rather than $\Lambda/Q$. 
Since kinematic restrictions affect large orders in perturbation 
theory, the ultraviolet behaviour of higher-twist operators and the entire 
higher-twist correction equally, the method outlined here will reproduce 
such parametric enhancements. Such enhancements will be seen to be 
crucial in understanding behaviour of power corrections to integrated 
fragmentation cross sections.

While the assumption of ultraviolet dominance of higher-twist corrections 
is definitively crude and clearly misses some features of higher-twist 
corrections (such as operators that do not mix into leading-twist ones 
through their ultraviolet behaviour), it should perhaps be measured by the 
simplicity with which some generic features of higher-twist corrections 
(and higher-order corrections in perturbation theory) can be incorporated.

\subsection{Calculating $A^i_{n,P}(x)$}

In this subsection we describe the calculation of the functions 
$A^i_{n,P}(x)$ in (\ref{withpower}). They can be obtained either 
from lowest-order Feynman diagrams 
with an infrared regulator or from infrared renormalons in the large-order 
behaviour of the coefficient functions $C_P^i$. The precise form of 
$A^i_{n,P}(x)$ depends on the method chosen, although some equivalences 
can be found \cite{BBZ}. We adhere to the second method. In this case, 
the functions $A^i_{n,P}(x)$ are defined as the residues of IR renormalon 
poles of the Borel transform of $C_P^i$. This is still not practical, because 
it would require us to calculate the perturbative expansion of $C_P^i$ to 
all orders. We therefore approximate the perturbative expansion by the 
series generated by inserting any number of fermion loops 
into the gluon line in diagrams that contribute at order $\alpha_s$. 

This class of diagrams falls into subclasses according to which parton 
is registered in the final state and where it originates from. 
Fig.~\ref{QCD96fig1} shows a contribution to the partonic cross 
section $d\sigma^q_P/d x$, where the registered quark originated at the 
primary virtual photon vertex. Another `primary' quark contribution is 
obtained if, in Fig.~\ref{QCD96fig1}, a gluon line instead of a fermion 
loop is cut. The registered quark can also come from a cut fermion 
loop. We refer to this contribution to $d\sigma^q_P/d x$, shown in 
Fig.~\ref{QCD96fig2}, as `secondary' 
quark contribution. Finally, the gluon cross section $d\sigma^g_P/d x$ 
is obtained from diagrams similar to that in Fig.~\ref{QCD96fig2}, 
but with a cut gluon line instead of a fermion loop. In addition, 
diagrams with counterterm insertions have to be included. 
The antiquark cross section is identical to the quark cross section. 

The set of diagrams with fermion loops is not really relevant by itself, 
as it typically yields small contributions to higher-order perturbative 
corrections. The main idea is to reinterpret the series in 
$(N_f\alpha_s)^n$ given by this set of diagrams as a series in 
$(\beta_0\alpha_s)^n$ and to restore
the full QCD $\beta$-function coefficient $\beta_0 =-1/(4\pi)[11 - 2 N_f/3]$ 
from the dependence on $N_f$ `by hand'. This substitution 
seems difficult to justify, but comparison with exact low-order results 
shows that it reproduces exact results quite well and that keeping 
corrections $(\beta_0\alpha_s)^n$ in higher orders resums important 
contributions. Further motivation can be found in \cite{BEN95,NEU95,DMW}. 
%%%%%%%%%%%%%%%%
% FIGURE 1
%%%%%%%%%%%%%%%%
\setlength{\unitlength}{0.7mm}
\begin{figure}[p]
\vspace{5.3cm}
%\hspace*{-3cm}
\begin{picture}(120,200)(0,1)
\mbox{\epsfxsize16.0cm\epsffile{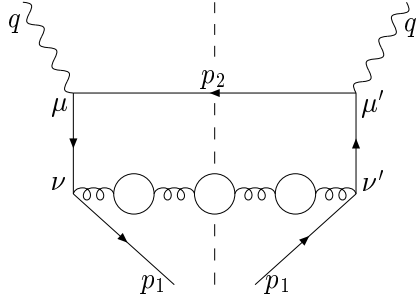}}
\end{picture}
\vspace*{-15.0cm}
\caption{\label{QCD96fig1}
Example of a `primary' quark contribution to the squared 
amplitude for $d\sigma_P^q/dx$. The set of all diagrams includes 
all attachments of the chain of fermion loops to the external quark 
line, the diagrams with a cut gluon line and an arbitrary number 
of fermion-loop insertions.}
\end{figure}
%%%%%%%%%%%%%%%%
% FIGURE 2
%%%%%%%%%%%%%%%%
\nopagebreak
\setlength{\unitlength}{0.7mm}
\begin{figure}[p]
\vspace{5.3cm}
%\hspace*{-3cm}
\begin{picture}(120,200)(0,1)
\mbox{\epsfxsize16.0cm\epsffile{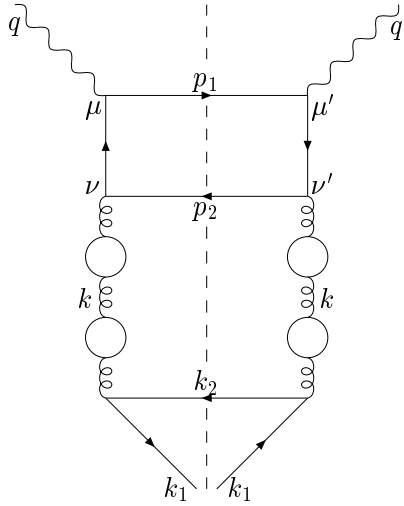}}
\end{picture}
\vspace*{-12.5cm}
\caption{\label{QCD96fig2}
Example of a `secondary' quark contribution to the squared 
amplitude for $d\sigma_P^q/dx$. The set of all diagrams includes 
all attachments of the chains of fermion loops to the external quark 
line and an arbitrary number 
of fermion-loop insertions.}
\end{figure}

The restoration of $\beta_0$ for the secondary quark contribution 
is especially delicate, since the fragmentation cross section 
depends on the `internal structure' of the fermion loop, that is 
the phase-space over $k_1$ and $k_2$ (see Fig.~\ref{QCD96fig2}) is 
non-trivially weighted. In such a situation, the association of fermion 
loops with running of the coupling is lost. To a certain extent, 
this instrinsic limitation of the method can be investigated by 
considering a fictitious theory with scalars rather than fermions in the 
fundamental representation. In this case we would compute scalar loop 
insertions and restore the full QCD $\beta$-function coefficient 
$\beta_0 =-1/(4\pi)[11 - N_s/6]$ from the dependence on $N_s$. 
We would like to obtain the same result 
as with fermions, since in both cases the important contributions 
are related to the non-abelian contribution to $\beta_0$. For observables 
that resolve the internal structure of fermion or scalar loops, however, 
the resulting functions $A_{n,P}^i(x)$ are not identical. This difference 
leads to some model dependence, which is analysed in Sect.~3.

The fermion-loop approximation (and the massive gluon calculation, 
discussed in Sect. 2.3) also neglects multiple 
gluon emission. This is an obvious shortcoming of the method 
relevant at small values of $x$. The resummation of $\ln x$ contributions 
is crucial for particle multiplicities and the shape 
of fragmentation functions in the small-$x$ region. 
In the following we consider the expansion in $Q^2$ at fixed $x$, leaving 
aside the question of the small-$x$ asymptotics at fixed $Q^2$.
The order of limits is probably important. Renormalon and 
small-$x$ resummations select different sets of diagrams and it is 
not known how to combine the two in a systematic way. See Sect.~4 for 
a further discussion of the problem. Related questions are 
also being discussed for Drell-Yan production \cite{KS95,BBDY}.

The evaluation of the two sets of diagrams shown in 
Fig.~\ref{QCD96fig1} and Fig.~\ref{QCD96fig2}, for an arbitrary 
number of fermion-loop insertions, and for their sum,
is straightforward by means of the dispersion technique 
discussed in Refs.~\cite{BEN95,NEU95}. For a moment, let us consider 
only diagrams with a cut fermion bubble. 
It is convenient to organize the calculation 
such that the integral over the gluon virtuality (invariant mass of 
the $q\bar{q}$ pair) is done last. 
The main object of our interest will be the distribution in the gluon 
virtuality $k^2$. To define it, consider for example the contribution of
the diagrams in Fig. 1 or 2 to $d \sigma_P/ d x$, summed over the number of 
fermion loops. It is given by\footnote{The following equation should be 
considered as schematic, because the limit $k^2\to 0$ needs some care. 
The correct treatment of this limit leads to (\ref{BS}) below. 
Alternatively, the equation can be understood as an expansion in 
$\alpha_s$.}
\begin{equation}
\frac{d \sigma_P^{[p,s]}}{d x} = 
\int \frac{d k^2}{k^2} \frac{\beta_0^f\alpha_s}
{\left| 1 + \Pi(k^2) \right|^2}~
\frac{d \sigma_P^{[p,s]}}{d x} \left(x, \frac{k^2}{Q^2}\right)~,
\label{defdistr}
\end{equation}
where $\Pi(k^2)$ is the $\overline{\mbox{MS}}$-renormalized fermion 
bubble, and the mass distribution can be written as
\begin{eqnarray}\label{xi-distributions}
\frac{1}{\sigma_0}\frac{d\sigma_P^{\rm{[p,s]}}}{dx}(x,\xi) &\equiv&  
\frac{8\pi}{N_c q^2}\int\! d\,{\rm Lips}[p_1,p_2,k]\,(2\pi)^4\,
\delta^{(4)}(q-p_1-p_2-k)
\frac{1}{k^2}{\cal M}_{\mu\nu}{\cal M}_{\mu'\nu'}^\ast
\nonumber\\
&&\hspace*{-1.5cm}\cdot\,\frac{2}{\beta_0^f}\,
\int\! d\,{\rm Lips}[k_1,k_2]\,(2\pi)^4\,
\delta^{(4)}(k-k_1-k_2)\,
\frac{N_f}{2}{\rm Tr}[\gamma_\nu\!\not\!k_1\gamma_{\nu'}\!\not\!k_2]
\,{\cal P}_{P;\,\mu\mu'}^{[p,s]}(x)~,
\end{eqnarray}
where $\xi = k^2/Q^2$.
In Eq.~(\ref{xi-distributions}), the notation for momenta and Lorentz 
indices corresponds to Fig.~2, and $\beta_0^f=N_f/(6\pi)$.
We denote by $[p]$ ($[s]$) the `primary' (`secondary')
quark contribution, while $\int\!d\,{\rm Lips[\ldots]}$ are 
Lorentz-invariant phase-space integrals, and 
${\cal M}_{\mu\nu}{\cal M}_{\mu'\nu'}^\ast$ is the matrix element 
for the primary $\gamma^*\to q\bar{q} g$ amplitude squared, divided
by the square of the quark electric charge. The 
projections are such that (in the $\gamma^*$ rest frame)
\begin{eqnarray}
&& {\cal P}_{L+T;\mu\mu'}^{[s]}(x) = -\frac{g_{\mu\mu'}}{4}\,
\delta\left(x-\frac{2 k_1 q}{q^2}\right)~,\\
&& {\cal P}_{L;\mu\mu'}^{[s]}(x) = \frac{k_{1,\mu}k_{1,\mu'}}
{4 |\vec k_1|^2}\,
\delta\left(x-\frac{2 k_1 q}{q^2}\right)~,
\end{eqnarray}
for the secondary contribution to the total and longitudinal fragmentation 
function. For the longitudinal projection we used $q_\mu{\cal M}_{\mu\nu}=0$,
and the primary quark contribution is obtained by replacing $k_1$ with $p_1$.  
The normalization factor $2/\beta_0^f$ in (\ref{xi-distributions})
may seem peculiar. It is chosen such that for inclusive quantities 
the distribution function coincides with the result that would be 
obtained from computing $\alpha_s$-corrections with a massive gluon.

Note that in the case of the primary quark contribution the phase-space
integral over $k_1,k_2$ is proportional to $k^2$, so that the result takes 
the form of the one-loop diagram calculated with a gluon of mass $k^2$. 
This ensures equivalence with the massive gluon calculation for this 
contribution. For the secondary quark contribution the projector 
depends on $k_1$ and modifies the phase-space integral. This is 
probably the simplest example of how event shapes in general are 
sensitive to the internal structure of fermion loops. 

Given the invariant mass distributions in $\xi$, the contributions of 
fermion-loop diagrams to higher-order perturbative corrections 
are obtained in terms of the logarithmic moment integrals \cite{BEN95}
\begin{equation}
\label{logmoments}
J_n(x) = \int_0^1 d\xi \,\ln^n\xi \,\frac{d}{d\xi}\frac{d\sigma_P^{[p,s]}}
{dx}(x,\xi) . 
\end{equation} 
The sum of the series, defined by a principal value prescription for 
the Borel integral, equals 
\begin{equation}\label{BS}
\frac{d\sigma_P^{(NNA)}}{dx} =
\int_0^1\!d\xi \,\Phi(\xi)\,\frac{d}{d\xi}\frac{d\sigma_P}{dx}(x,\xi) 
+\left[\frac{d\sigma_P}{dx}(x,\xi_L)-
\frac{d\sigma_P}{dx}(x,0)\right] ,
\end{equation}
where $\xi_L < 0$ is the position of the Landau pole in the
strong coupling and the function $\Phi(\xi)$ is specified 
in Eq.~(2.25) of the second reference in \cite{BEN95}.
We shall make use of these results in Sect.~5. 

To obtain the functions $A^i_{n,P}(x)$, it is not necessary to perform 
the final integration over $k^2$ ($\xi$). The infrared renormalon 
residues can be read off as coefficients of non-analytic terms in the
expansion of the $k^2$ distribution $d\sigma_P/dx (x,\xi)$ at 
small $\xi$ \cite{BBZ,BEN95} 
\begin{equation}\label{expansion}
\frac{d\sigma_P}{dx}(x,\xi) = \frac{d\sigma_P}{dx}(x,\xi\to0)
+\frac{C_F\alpha_s}{2\pi}\left\{A_{1,P}(x)\,\sqrt{\xi} +
A_{2,P}(x)\,\xi\ln\xi + A_{4,P}(x)\,\xi^2\ln\xi + \ldots\right\}. 
\end{equation}
Eq.~(\ref{expansion}) thus supplies all information required for the 
convolutions in (\ref{withpower}).

Up to this point, we considered the  partonic 
fragmentation cross sections $H=q,g$ in (\ref{DGLAP}). To extract 
the  coefficient functions $C_P^i$, the partonic 
fragmentation functions $D_i^{q,g}$ have to be calculated and subtracted 
in the same approximation. The need for subtractions is also 
reflected in the fact that the limit $\xi\to 0$ in (\ref{expansion}) does 
not exist, because $d\sigma_P/dx\sim\ln\xi$ at small $\xi$. As a 
consequence, the integrals in (\ref{logmoments}) and (\ref{BS}) also 
diverge at small $\xi$. Once the partonic fragmentation functions 
are computed, the subtractions can be implemented into these 
equations by generalizing the case of ultraviolet subtractions 
discussed in \cite{BEN95} to infrared subtractions.

We emphasize that the power corrections added in (\ref{withpower})
(that is the functions $A_{n,P}(x)$) depend on the factorization 
scheme and, for instance in the `annihilation scheme', 
would differ from those in the $\overline{\mbox{MS}}$-scheme, because 
the definition of the leading-twist fragmentation function can 
include an arbitrary set of power corrections. We will be working 
in the $\overline{\mbox{MS}}$-scheme. In this particular scheme, we
do not need to perform the subtractions explicitly and 
the functions $A_{n,P}(x)$ can already be obtained from the 
partonic cross sections. To see this, let us sketch how
collinear factorization is performed for the set of diagrams considered above.

To solve (\ref{DGLAP}) for $C_P^i$ it is convenient to count powers 
of $N_f$ for a given diagram, where $N_f$ denotes the 
number of flavours. We expand the partonic fragmentation functions
\begin{equation}
\label{dexp}
D_{i}^k(x) = D_i^{k,[0]}(x) + \frac{1}{N_f}\,D_i^{k,[1]}(x) + \ldots~,
\end{equation}
where $D_i^{k,[n]}(x)$ are power series in $N_f\alpha_s$.
The leading order (in $1/N_f$) contributions are
\begin{eqnarray}
D_q^{q,[0]}(x) &=& \delta(1-x)~,\\
D_g^{g,[0]}(x) &=& \delta(1-x)\,\frac{1}{1-\beta_0^f\alpha_s/\epsilon}~,\\
D_q^{g,[0]}(x) &=& 0~,\\
D_g^{q,[0]}(x) &=& \frac{\alpha_s}{2\pi\epsilon} P_{g\to q}(x) + 
{\cal O}(N_f^2\alpha_s^2)~,
\end{eqnarray}
where $P_{g\to q}(x) = N_f/2\,[x^2+(1-x)^2]$ denotes the DGLAP splitting 
function, $\beta_0^f= N_f/(6\pi)$ is the fermionic contribution to the 
$\beta$ function, and $\epsilon = (4 - d)/2$. 
Note that $D_g^{q,[0]}(x)$ contains an entire 
series in $N_f\alpha_s$, of which we show the first term only.
The factor $1/(1-\beta_0^f\alpha_s/\epsilon)$ in $D_g^{g,[0]}(x)$ comes 
from counterterm insertions in fermion loops.
 Since the Born diagram leads to a non-vanishing coefficient 
function only for $C_T^q(x)=\delta(1-x)$, we find for the 
coefficient functions, in the fermion-loop approximation,
\begin{eqnarray}
\label{gluoncoeff}
C_L^g(x) &=& \left(1-\frac{\beta_0^f\alpha_s}{\epsilon}\right)\,
\frac{d\sigma_L^g}{dx} = \frac{C_F\alpha_s}{2\pi}\,\frac{4 (1-x)}{x}~,\\
\label{sub}
C_L^q(x) &=& \frac{d\sigma_L^q}{dx} - C_L^g \ast D_g^{q,[0]}~,\\
C_T^g(x) &=& \left(1-\frac{\beta_0^f\alpha_s}{\epsilon}\right)
\left(\frac{d\sigma_T^g}{dx} - D_q^{g,[1]}\right)~,\\
C_T^q(x) &=& \frac{d\sigma_T^q}{dx} - D_q^{q,[1]} - C_T^g \ast D_g^{q,[0]}~.
\end{eqnarray}
The asterisk denotes convolution. The coefficient functions have finite 
limits as $\epsilon\to 0$. Note that
in the fermion-loop approximation, the longitudinal 
gluon coefficient function has no corrections beyond first order.
The graphs with uncut fermion loops are cancelled by analogous 
contributions to $D_g^g$. 
On the other hand, the longitudinal quark coefficient function 
and the transverse coefficient functions 
involve non-trivial subtractions.

However, as long as we are interested only in power corrections 
(infrared renormalons) 
and not in the complete higher-order perturbative corrections, 
we do not need to calculate these subtractions explicitly.  
The partonic fragmentation 
functions needed in our approximation have convergent series 
expansions and do not contribute to infrared renormalon residues. 
This property is specific to minimal subtraction schemes, where 
partonic fragmentation functions are expressed as pure poles in $\epsilon$. 
For the $\xi$ distributions introduced above, this implies that 
the $\ln\xi$ terms are removed, but all other non-analytic terms,  
and in particular the functions $A_P^i(x)$,  
remain unaltered by factorizing the fragmentation functions. 

It is important that the mixing between quark and gluon operators 
at leading twist repeats itself at the level of power corrections 
and renormalons. Although the gluon coefficient functions are convergent 
series in the fermion-loop approximation, this does not mean that 
gluon fragmentation does not give rise to power corrections. 
Because of mixing, the infrared renormalons in quark coefficient 
functions due to the secondary quark contribution are cancelled by 
ultraviolet renormalons in quark matrix elements of gluon operators of 
higher twist. The same gluon operators also have non-vanishing 
gluon matrix elements, which are set to zero only in the formal 
large-$N_f$ limit.

\subsection{Massive gluon scheme}

It can be shown \cite{BBZ,BEN95} that 
for sufficiently inclusive observables, the functions $A^i_{n,P}(x)$ 
obtained from the residues of IR renormalons in the above 
approximation to the perturbative series can also be obtained from 
the $\alpha_s$ corrections alone, provided they are computed with 
non-vanishing gluon mass. In this case, these functions can be read off 
from non-analytic terms in the small-mass expansion.\footnote{The 
restriction to non-analytic terms is clear. Analytic terms in 
$\lambda^2$ also arise from large $k^2$, where the propagator 
$1/(k^2+\lambda^2)$ can be Taylor-expanded in $\lambda^2$.} This 
equivalence does not hold for general event shapes, which 
are sensitive to the internal structure of fermion loops, as is the case for 
the secondary quark contribution in Fig.~\ref{QCD96fig2} (and as discussed
in Ref.~\cite{NAS95}). 

Although in this situation a formal justification of the massive 
gluon scheme is lacking, except, perhaps, as an ad hoc implementation 
of the general idea of scale separation by an explicit infrared cut-off, 
we find it useful to check whether the massive gluon calculation 
reproduces the gross features of the fermion-loop calculation. If so, 
one could take advantage of the relative simplicity of the massive 
gluon calculation. 

With this motivation in mind, we will present also the distribution functions 
in $\xi$ in the massive gluon scheme (see also \cite{DAS96}). 
We will find that the comparison between the fermion loop and massive 
gluon calculation has to be done case by case, with varying conclusions. 

\subsection{Summary of distribution functions}

In this subsection we present our results for the distributions in $\xi$
corresponding to the detection of the primary and the secondary quark, as 
well as those corresponding to the detection of a gluon, calculated in the
massive gluon scheme. We give the exact expressions, as well as the
small-$\xi$ expansions corresponding to (\ref{expansion}), and we extract
the coefficients $A^i_{j,P}(x)$ of the power corrections.
As explained above, the gluon coefficient function is trivial in the 
large-order approximation we have adopted and does not indicate power 
corrections. The role of the gluon coefficient is taken by the 
secondary quark contribution, Fig.~\ref{QCD96fig2}, to the 
quark coefficient function. The results presented below refer to the 
sum of quark and antiquark contributions, which amounts to multiplying 
the quark contribution by two.

\subsubsection{Primary quark contribution}

For the primary quark contribution to $d\sigma_P^q/dx$ (Fig.~\ref{QCD96fig1}),
the distribution in $k^2$ coincides with the partonic quark 
fragmentation cross section calculated with a massive gluon. Defining 
$\xi=k^2/Q^2$ as before we have
\begin{eqnarray}
\frac{1}{\sigma_0}\frac{d\sigma_L^{q,[p]}}{dx} &=& \frac{C_F\alpha_s}{2\pi}
\cdot 2\cdot\Theta(1-\xi-x)
\Bigg[1+\xi \left(\frac{6}{x}-2-\frac{2}{1-x}\right)+\xi^2
\bigg(-\frac{6}{x}-\frac{4}{1-x}
\nonumber\\
&&{}+\frac{1}{(1-x)^2}\bigg)
+\frac{2\xi}{x^2}(2x+3\xi)\ln\frac{\xi}{(1-x)(x+\xi)}\Bigg]~,
\end{eqnarray}
\begin{eqnarray}
\frac{1}{\sigma_0}
\frac{d\sigma_T^{q,[p]}}{dx} &=& \frac{C_F\alpha_s}{2\pi}\cdot 2\cdot
\Bigg\{\delta(1-x)\Bigg[2(1+\xi)^2\left({\rm Li}_2(-\xi)-\frac{1}{2}\ln^2\xi
+\ln \xi \ln(1+\xi)+\frac{\pi^2}{6}\right)
\nonumber\\
&&{}-(3+2\xi)\ln \xi -\frac{7}{2}-2\xi
\Bigg] +\Theta(1-\xi-x)\Bigg[-\frac{3-5\xi^2}{2(1-x)}+\frac{\xi}{(1-x)^2}
\nonumber\\
&&{}+\frac{\xi^2}{2(1-x)^3}
+\frac{\xi}{x+\xi}-\frac{6\xi(1-\xi)}{x}-\frac{1-x}{2}+3\xi
\nonumber\\
&&{}
-\Bigg(\frac{2(1+\xi)^2}{1-x}-1-x-2\xi+\frac{4\xi}{x}
+\frac{6\xi^2}{x^2}\Bigg)\ln\frac{\xi}{(1-x)(x+\xi)}\Bigg]\Bigg\}.
\end{eqnarray}
The expansion in $\xi$ gives
\begin{eqnarray}
\frac{1}{\sigma_0}
\frac{d\sigma_L^{q,[p]}}{dx} &=& \frac{C_F\alpha_s}{2\pi}\cdot 2\cdot
\Bigg[ 1+\xi\ln \xi\,A_{2,L}^{q,[p]}(x) 
+\xi\left(\frac{6}{x}-2-\frac{2}{[1-x]_+}+\frac{4}{x}\ln\frac{1}{x(1-x)}
\right)
\nonumber\\
&&{}
+\xi^2 \ln \xi \,A_{4,L}^{q,[p]}(x) + O(\xi^2)
\Bigg],
\end{eqnarray}
where
\begin{eqnarray}
A_{2,L}^{q,[p]}(x) &=& 2\delta(1-x)+\frac{4}{x},\\
A_{4,L}^{q,[p]}(x) &=& \delta'(1-x)+4\delta(1-x)+\frac{6}{x^2}.
\end{eqnarray}
The $\delta$ functions arise from expanding singular functions around 
the phase-space boundary $1-x-\xi$. For the transverse cross section, 
we keep only non-analytic terms in the expansion; then
\begin{eqnarray}
\frac{1}{\sigma_0}
\frac{d\sigma_T^q}{dx} &=& \frac{C_F\alpha_s}{2\pi}\cdot 2\cdot
\Bigg[-\ln\xi\left(\frac{1+x^2}{[1-x]_+}+\frac{3}{2}\delta(1-x)\right)
+\xi \ln \xi \,A_{2,T}^{q,[p]}(x)\nonumber\\
&&{}
+\xi^2 \ln \xi \,A_{4,T}^{q,[p]}(x)+O(\xi^3\ln\xi)\Bigg],
\end{eqnarray}
where
\begin{eqnarray}
A_{2,T}^{q,[p]}(x) &=& -\frac{4}{[1-x]_+}+2-\frac{4}{x}-2\delta(1-x)
+\delta'(1-x)~,\\
A_{4,T}^{q,[p]}(x) &=& -\frac{2}{[1-x]_+}-\frac{6}{x^2}-\frac{5}{2}\delta(1-x)
-\frac{1}{4}\delta''(1-x)~.
\end{eqnarray}

\subsubsection{Secondary quark contribution}

Because of cancellations between the longitudinal and tranverse 
cross sections it is more convenient to quote the distribution in $\xi$ 
for the longitudinal contribution and the sum of transverse and 
longitudinal contributions,
\begin{eqnarray}
\frac{1}{\sigma_0}\frac{d \sigma_L^{q,[s]}}{d x} &=&  
\frac{C_F\alpha_s}{2\pi}\cdot 2\cdot \Theta(1-x) \Theta(x-\xi)
\Bigg[-\frac{15 \xi (1 + \xi)}{8 x^2} 
\nonumber \\
&& + \,\frac{3 (1 - \xi)(5 \xi^2 + 10 \xi x^2 - 3 x^4)}{
16 \xi x^2} \ln \xi  - 6 \frac{\xi}{x} \ln x \ln \frac{x}{\xi}
\nonumber \\
&& + \,\frac{3}{8\xi x} \left(5 \xi + 2 \xi^2 + 5 \xi^3 + 6 \xi x +
6 \xi^2 x - 3 x^2 - 14 \xi x^2 - 3 \xi^2 x^2 + 3 x^3 +
3 \xi x^3\right)
\nonumber \\
&& +\,\frac{3}{16\xi x} \left(5 \xi + 18 \xi^2 + 5 \xi^3 - 16 \xi x -
16 \xi^2 x + 3 x^2 - 2\xi x^2 + 3 \xi^2 x^2\right)
\ln \frac{\xi}{x^2}
\nonumber\\
&& -\,\frac{3}{8 \xi x^2} 
\Big(5 \xi^2 + 18 \xi^3 + 
5 \xi^4 - 6 \xi x^2 + 4 \xi^2 x^2 - 6 \xi^3 x^2 - 
3 x^4 + 2 \xi x^4 
\nonumber\\
&& - \,3 \xi^2 x^4 \Big) \,T(\sqrt{\xi},x)\Bigg]~,
\label{dsigmaldx} 
\end{eqnarray}
\begin{eqnarray}
\frac{1}{\sigma_0}\frac{d \sigma_{L+T}^{q,[s]}}{d x} &=&  
\frac{C_F\alpha_s}{2\pi}\cdot 2\cdot \Theta(1-x) \Theta(x-\xi)
\Bigg[\frac{36 \,(1 + \xi)}{x^2} 
\nonumber\\
&& - \,\frac{12}{\xi x^3} \left(4 \xi^2 - x^2 + 2 \xi x^2 - 
\xi^2 x^2 + x^3 + \xi x^3\right)
+ \frac{6 \,(1 - \xi)}{\xi x^2}\,\left(x^2+3\xi\right)\,\ln \xi
\nonumber\\
&& -\,\frac{6}{\xi x^3}\,\left(- 2 \xi^2 - 2 \xi x - 2 \xi^2 x + x^2 + 
4 \xi x^2 + \xi^2 x^2\right) \,\ln\frac{\xi}{x^2}
\nonumber\\
&& - \, \frac{24 \,(1 + x+\xi)}{x^2} \,\ln x\ln\frac{x}{\xi} 
- \, \frac{12}{\xi x^2}\,(1 + \xi)^2 \,\left(x^2+3 \xi
\right)\,T(\sqrt{\xi},x)\Bigg]~,
\label{dsigmatotdx} 
\end{eqnarray}
where
\begin{eqnarray}
T(\lambda,x) & \equiv & 
\int_{1/x}^1 \!d t \,\frac{\ln (\lambda t)}{1 + \lambda^2 t^2} = 
\frac{{\rm i}}{2 \lambda} \left[\ln \lambda 
\ln \frac{1 - {\rm i} \lambda}{1 + {\rm i} \lambda} - 
\ln \frac{\lambda}{x} \ln \frac{x - {\rm i} \lambda}{x + 
{\rm i} \lambda} \right.
\nonumber\\
&& +\,\left. {\rm Li}_2 ({\rm i} \lambda) - {\rm Li}_2 (- {\rm i} \lambda)
- \left({\rm Li}_2 \left({\rm i} \frac{\lambda}{x}\right) - 
{\rm Li}_2 \left(- {\rm i} \frac{\lambda}{x}\right) \right) \right]~.
\label{litx} 
\end{eqnarray} 
The $\xi\to 0$ limits and the first non-analytic terms in the expansion 
at small $\xi$ are given by
\begin{eqnarray}
\frac{1}{\sigma_0}\frac{d\sigma_L^{q,[s]}}{dx} 
&=&\frac{C_F\alpha_s}{2\pi}\cdot 2\cdot \Bigg[
\frac{4}{x} -6 x+2 x^2+6 \ln x +\xi \ln \xi \,A_{2,L}^{q,[s]}(x) 
\nonumber\\
&& \,+\xi^2 \ln \xi \,A_{4,L}^{q,[s]}(x) + \ldots\Bigg],
\label{secl}\\
\frac{1}{\sigma_0}\frac{d\sigma_{L+T}^{q,[s]}}{dx} 
&=&\frac{C_F\alpha_s}{2\pi}\cdot 2\cdot \Bigg[-\ln\xi
\Big(2\cdot \frac{3}{N_f}\cdot\frac{1}{C_F}
\cdot \left[P_{q\to g}\ast P_{g\to q}\right](x)\Big)
+\xi \ln \xi \,A_{2,L+T}^{q,[s]}(x)
\nonumber\\
&& \,+\xi^2 \ln \xi \,A_{4,L+T}^{q,[s]}(x) + \ldots\Bigg],
\label{sectot}
\end{eqnarray}
where
\begin{eqnarray}
A_{2,L}^{q,[s]}(x) &=& \frac{6}{5x^3}+\frac{4}{x}-6+\frac{4}{5}x^2+
\frac{6\ln x}{x}~,
\label{a2ls}
\\
A_{4,L}^{q,[s]}(x) &=& -\frac{16}{35x^5}+\frac{16}{5x^3}-\frac{4}{x^2}
+\frac{8}{5}-\frac{12}{35}x^2~,
\\
A_{2,L+T}^{q,[s]}(x) &=& \frac{6}{5x^3}-\frac{11}{x}+9+\frac{4}{5}x^2-
6\ln x~,
\label{a2lts}
\\
A_{4,L+T}^{q,[s]}(x) &=& -\frac{24}{35x^5}+\frac{12}{5x^3}-\frac{4}{x}
+\frac{12}{5}-\frac{4}{35}x^2~.
\label{a4lts}
\end{eqnarray}
Note that the expansions in $\xi$ are valid only for $x>\sqrt{\xi}$, 
although $x$ can be as small as $\xi$. The minimal invariant mass 
of the $q\bar{q}$ pair, however, is attained at $x=\sqrt{\xi}$. 

The $\xi\to 0$ limits in (\ref{secl}) and (\ref{sectot}) require 
explanation. First note that although the diagrams in Fig.~\ref{QCD96fig1} 
and Fig.~\ref{QCD96fig2} start at order $\alpha_s^2$, our resulting 
distribution functions are written as order $\alpha_s$. Technically, 
this can be best understood by considering the Borel transform 
of the series generated by these diagrams. Since there is no diagram 
at order $\alpha_s$, one would expect the Borel transform to vanish 
linearly in $u$, the Borel parameter, at small $u$. However, the 
$g\to q\bar{q}$ splitting amplitude contains a collinear divergence, 
which manifests itself as a pole $1/u$. Consequently, the Borel transform 
approaches a constant for small $u$, which corresponds to an $\alpha_s$ 
contribution. The collinear pole in $u$ indicates that collinear 
subtractions need to be performed. Indeed, if $C_L^g\ast D_g^{q,[0]}$ 
is subtracted as required by (\ref{sub}), the constant term in the 
Borel transform vanishes and the secondary quark contribution to the 
quark coefficient function is of order $\alpha_s^2$, as it should be. 

To check this cancellation, we observe that the leading term in 
(\ref{secl}) can be rewritten as 
\begin{equation}
\label{forty}
\frac{C_F\alpha_s}{2\pi}\cdot 2\cdot \Bigg[
\frac{4}{x} -6 x+2 x^2+6 \ln x\Bigg] = 2\cdot\frac{3}{N_f}\cdot
\left[C_L^g\ast P_{g\to q}\right](x),
\end{equation}
with $C_L^g$ defined in (\ref{gluoncoeff}). As expected from the 
above discussion, the $\xi\to0$ limit of the secondary quark contribution 
is just the gluon coefficient function convoluted with collinear splitting 
into a $q\bar{q}$ pair. The factor 2 accounts for the sum over 
quark and antiquarks. The factor $3/N_f$ is the large-$N_f$ limit of 
$\alpha_s/(2\pi)\cdot\ln Q^2/\Lambda^2$ associated with the 
$g\to q\bar{q}$ amplitude.

The $\ln\xi$ term in the sum of the longitudinal and transverse cross 
sections can likewise be interpreted as a convolution of two 
subsequent $q (\bar{q})\to g$ and $g\to q (\bar{q})$ splittings, as 
indicated in (\ref{sectot}). 

We should stress again that these complications associated with collinear 
factorization do not affect the non-analytic terms in the expansion in 
$\xi$, except for $\ln \xi$, and therefore do not affect the expressions 
for $A_{n,P}^i$. Note that for heavy quark fragmentation, the secondary 
quark contribution would not give rise to power corrections, because 
the invariant mass distribution is cut off at $4 m_Q^2$.

\subsubsection{Massive gluon scheme}

For completeness and later comparison, we quote the distribution 
functions in the massive gluon scheme. The distribution functions are 
given as the partonic fragmentation cross sections to order $\alpha_s$, 
computed with a massive gluon. The quark fragmentation cross sections
are identical to our primary quark contributions, Sect. 3.1.1. 
The gluon cross sections read
\begin{eqnarray}
\frac{1}{\sigma_0}\frac{d\sigma_L^{g,{\rm MG}}}{dx} &=&
\frac{C_F\alpha_s}{2\pi}\Theta(1+\xi-x)\Theta(x-2\sqrt{\xi})
\nonumber\\
&&\cdot \,4\,(1+\xi)\frac{1+\xi-x}{\sqrt{x^2-4\xi}}
\left[1-\frac{2\xi}{x}\frac{1}{\sqrt{x^2-4\xi}}
\ln\frac{(x+\sqrt{x^2-4\xi})^2}{4\xi}\right]~,
\label{lmg}
\end{eqnarray}
\begin{eqnarray}
\frac{1}{\sigma_0}\frac{d\sigma_{L+T}^{g,{\rm MG}}}{dx} &=& 
\frac{C_F\alpha_s}{2\pi}\Theta(1+\xi-x)
\Theta(x-2\sqrt{\xi})
\Bigg[-4\sqrt{x^2-4\xi}
\nonumber\\&&{}+
\frac{2}{x}[2(1+\xi)(1+\xi-x)+x^2]
\ln\frac{(x+\sqrt{x^2-4\xi})^2}{4\xi}\Bigg]~.
\label{ltmg}
\end{eqnarray}
As $\xi\to 0$, the longitudinal cross section reduces to 
(\ref{gluoncoeff}). Expansion in $\xi$ results in
\begin{eqnarray}
\frac{1}{\sigma_0}\frac{d\sigma_L^{g,{\rm GM}}}{dx} &=& 
\frac{C_F\alpha_s}{2\pi}
\Bigg[\frac{4(1-x)}{x} + \xi\ln\xi \,A_{2,L}^{g,\rm MG}(x) + \xi^2\ln\xi
\,A_{4,L}^{g,\rm MG}(x) +\ldots\Bigg]
\\
\frac{1}{\sigma_0}\frac{d\sigma_{L+T}^{g,{\rm GM}}}{dx} &=& 
\frac{C_F\alpha_s}{2\pi}
\Bigg[-2\cdot\ln\xi\,\frac{(1+(1-x)^2)}{x} + \xi\ln\xi 
\,A_{2,L+T}^{g,\rm MG}(x) 
\nonumber\\
&&  + \,\xi^2\ln\xi\,A_{4,L+T}^{g,\rm MG}(x) +\ldots\Bigg],
\end{eqnarray}
where again we have kept only the leading term and the first few 
non-analytic terms. The coefficients of power corrections are now given by
\begin{eqnarray}
A_{2,L}^{g,\rm MG}(x) &=& \frac{8(1-x)}{x^3}~, 
\label{a2gmg}
\\
A_{4,L}^{g,\rm MG}(x) &=& \frac{8(2-x)}{x^3}+\frac{32(1-x)}{x^5}~,
\\
A_{2,L+T}^{g,\rm MG}(x) &=& - \frac{8}{x}+4-2\delta(1-x)~,
\label{a2ltmg}
\\
A_{4,L+T}^{g,\rm MG}(x) &=& -\frac{4}{x}-3\delta(1-x)-\delta'(1-x)~.
\end{eqnarray}
Our results in the massive gluon scheme coincide with those 
obtained by Dasgupta and Webber \cite{DAS96}.

%%%%%%%%%%%%%%%%%%%%%%%%%%%%%%%%%
% SECTION 3
%%%%%%%%%%%%%%%%%%%%%%%%%%%%%%%%%

\section{Power corrections to fragmentation functions}

In this section we analyse the $x$ dependence of power corrections. 
We present predictions in the fermion-loop approximation and then discuss 
in detail the ambiguities related to the restoration of the gluon 
fragmentation component. To this end, we compare the 
fermion-loop calculation with the massive gluon calculation and 
check the effect of replacing fermions by scalars. Once again, `massive gluon 
calculation' refers to the identification of power corrections through 
non-analytic contributions in the expansion of order-$\alpha_s$ 
corrections in a small gluon mass, as done in Ref.~\cite{DAS96}.
Finally, in Sect.~3.5, we abstract from the discussion the generic 
dependences on $x$ and formulate a simple phenomenological parametrization 
of $1/Q^2$ corrections consistent with our results.

\subsection{Results}

To illustrate the magnitude of the leading $1/Q^2$ power corrections to the 
fragmentation cross sections in $e^+ e^-$ annihilation in the 
fermion-loop approximation, we rewrite 
(\ref{withpower}) as
\begin{equation} 
\label{newpower}
\frac{d\sigma_P}{dx}(x,Q^2) = F_P(x,Q^2)\left[1 + 
H_{2,P}(x,Q^2)\,\frac{\Lambda^2}{Q^2} + \ldots\right],
\end{equation}
where
\begin{equation}
F_P(x,Q^2) = \sum_i \left[C_P^i\ast D_i\right](x,Q^2) = 
\sum_i\int_x^1\frac{dz}{z}\,C_P^i(z,Q^2/\mu^2)\,D_i(x/z,\mu)
\end{equation}
denotes the leading-twist cross section.
The power correction is given by
\begin{equation}
H_{2,P}(x,Q^2)\equiv \sum_i K_2^i\,\frac{\left[A_{2,P}^i\ast D_i\right](x,Q^2)}
{F_P(x,Q^2)} = 
\frac{1}{F_P(x,Q^2)}\sum_i\int_x^1\frac{dz}{z}\,K_2^i\,A_{2,P}^i(z)
\,D_i(x/z,\mu).
\end{equation}
In the following, we set $K_2^q=1$. Since in the fermion-loop approximation 
we only have a quark (antiquark) contribution, no information is lost, as the 
over-all scale is then set by $\Lambda^2$ in (\ref{newpower}). We use 
the ALEPH parametrization \cite{ALEPH} of the (light) quark (and later, 
gluon) fragmentation function at $Q=\sqrt{s}=22\,$GeV and their value 
$\alpha_s(22 {\rm GeV})= 0.164$. In evaluating $F_P(x,Q^2)$, we use the 
lowest-order approximation to the leading-twist coefficient function 
$C_P^q$. 

%%%%%%%%%%%%%%%%
% FIGURE 3
%%%%%%%%%%%%%%%%
\begin{figure}[p]
%   \vspace{-4cm}
%   \epsfysize= 7.5cm
%   \epsfxsize=10.2cm
   \centerline{\epsffile{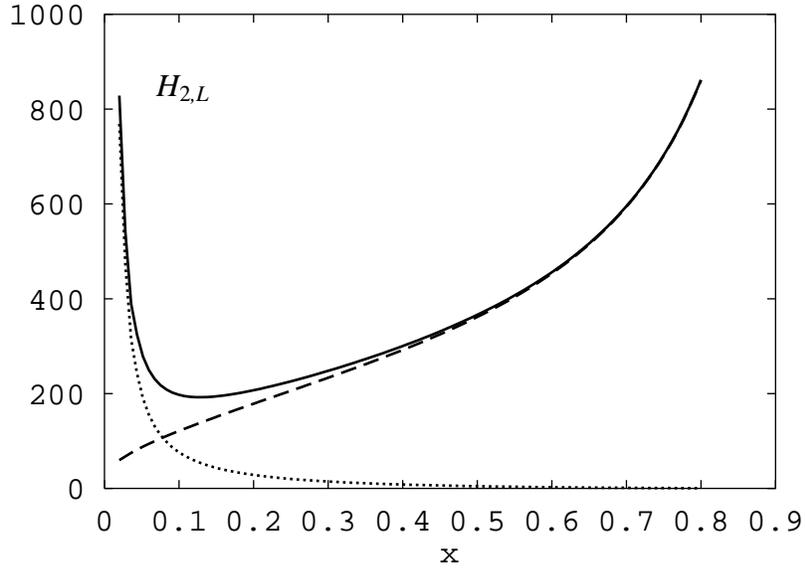}}
%   \vspace*{-15cm}
\caption{\label{figLong} Shape of $1/Q^2$ power correction $H_{2,L}(x)$ 
to the longitudinal fragmentation cross section. Dashed line: primary 
quark contribution. Dotted Line: secondary quark contribution. 
Solid line: sum of both. 
}
\end{figure}
%%%%%%%%%%%%%%%%
% FIGURE 4
%%%%%%%%%%%%%%%%
\begin{figure}[p]
%   \vspace{-4cm}
%   \epsfysize=7.5cm
%   \epsfxsize=10.2cm
   \centerline{\epsffile{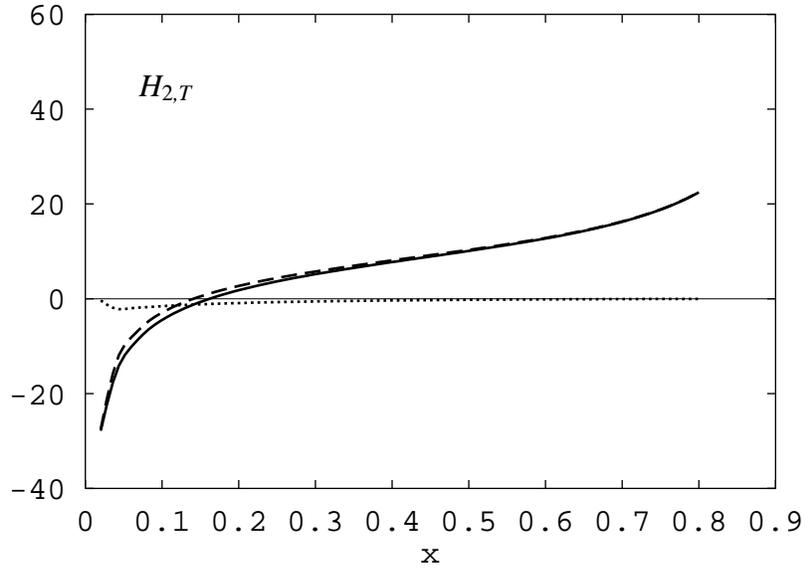}}
%   \vspace*{-15cm}
\caption{\label{figTrans} Shape of $1/Q^2$ power correction $H_{2,T}(x)$ 
to the transverse fragmentation cross section. Dashed line: primary 
quark contribution. Dotted Line: secondary quark contribution. 
Solid line: sum of both. 
}
\end{figure}
%%%%%%%%%%%%%%%%
% FIGURE 5
%%%%%%%%%%%%%%%%
\begin{figure}[t]
%   \vspace{-4cm}
%   \epsfysize=25.2cm
%   \epsfxsize=18cm
   \centerline{\epsffile{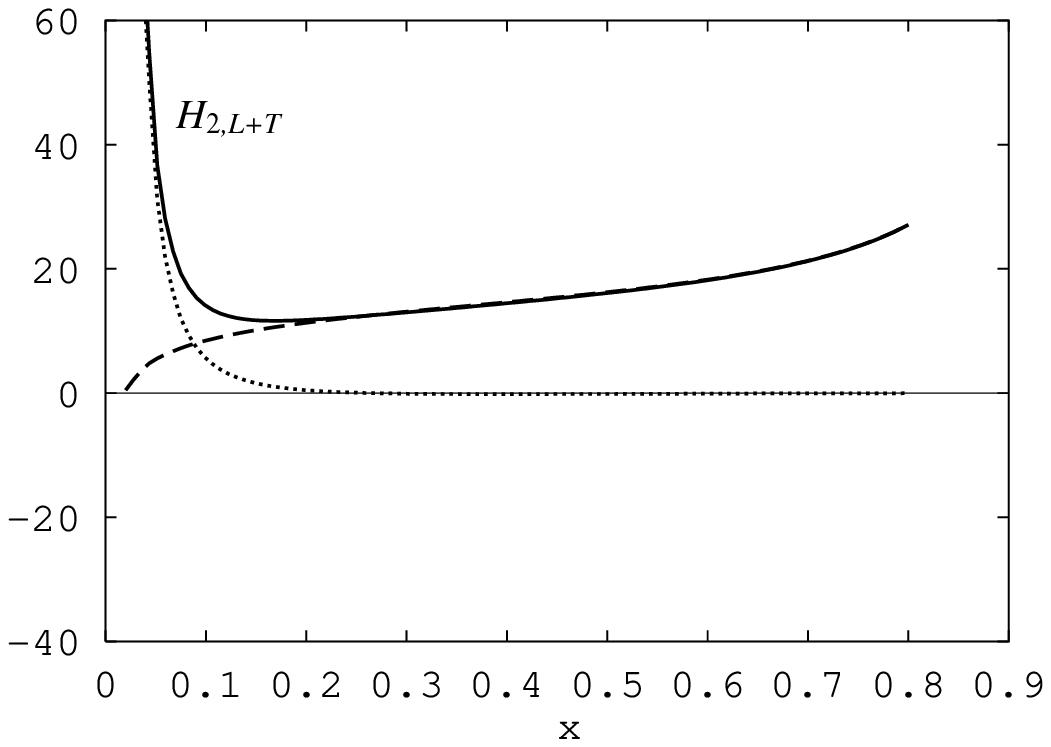}}
%   \vspace*{-15cm}
\caption{\label{figTotal} Shape of $1/Q^2$ power correction $H_{2,L+T}(x)$ 
to the sum of longitudinal and transverse 
fragmentation cross sections. Dashed line: primary 
quark contribution. Dotted Line: secondary quark contribution. 
Solid line: sum of both. 
}
\end{figure}

The result for $H_{2,P}(x)$ for the longitudinal and transverse fragmentation 
cross sections, summed over all hadrons, as a function of energy fraction 
$x$ is shown in Figs.~\ref{figLong} and \ref{figTrans}, respectively. The 
sum of longitudinal and transverse cross sections, which differs from the 
total cross section by the small asymmetric contribution due to 
$\gamma/Z^0$ interference, is shown in Fig.~\ref{figTotal}. 
We recall that $F_L/F_T \sim\alpha_s$, so that the relative 
magnitude of power corrections is enhanced for longitudinal 
fragmentation as reflected by the scales on the vertical axes in 
Fig.~\ref{figLong} and \ref{figTrans}. 
We note that for $x > 0.2$ the power correction to 
both, longitudinal and transverse, cross sections is dominated by 
primary quark fragmentation. Such a qualitative behaviour is expected, 
since the average energy of a quark connected to the `hard' $\gamma^*$ or 
$Z^0$ vertex is much larger than the average energy of a quark 
originating from gluon splitting. Since for the primary quark 
contribution the restoration of $\beta_0$ from the dependence on $N_f$ 
is unproblematic and the fermion-loop and massive gluon calculations 
coincide, the method yields an unambiguous prediction at  $x > 0.2$. 
At smaller values of $x$ the secondary quark contribution becomes 
important for the longitudinal cross section and for the 
sum of longitudinal and transverse 
cross sections, but remains small for the transverse one. 
Comparing Figs.~\ref{figLong}--\ref{figTotal} with the corresponding 
figures in Ref.~\cite{DAS96}, obtained from the massive gluon calculation, 
one concludes that the results are qualitatively similar for the 
longitudinal cross section, but differ drastically at small $x$ for the 
others. The difference can be traced to the absence of a $1/x^3$ term 
in $A_{2,L+T}^{g,\rm MG}(x)$, Eq.~(\ref{a2ltmg}), as compared to 
$A_{2,L+T}^{q,[s]}(x)$, Eq.~(\ref{a2lts}), and will be discussed 
in detail in Sect.~3.4.

Let us comment on the normalization of $H_{2,P}(x)$. With $K_2^i=1$, 
the overall normalization of $A_{2,P}$ was obtained by setting 
$C_F\alpha_s/(2 \pi)\,\xi\ln\xi\to \Lambda^2/Q^2$ in the 
distribution functions. If we took the normalization literally from 
the IR renormalon ambiguity of the Borel integral, divided by $\pi$, 
we would have to substitute
\begin{equation}
\frac{C_F\alpha_s}{2 \pi}\,\xi\ln\xi\,\to \,
\frac{C_F e^{5/3}}{2\pi (-\beta_0)}
\,\frac{\Lambda^2}{Q^2} \approx 1.6\,\frac{\Lambda^2}{Q^2},
\end{equation}
where $\Lambda=Q\,\exp(1/(2\beta_0\alpha_s(Q)))$. With this 
substitution, using $\alpha_s$ as given above 
and $H_{2,L+T}(x)\sim 15$ for intermediate $x$, the 
power correction in Fig.~\ref{figTotal} scales as 
$(0.7\,\mbox{GeV}/Q)^2$, very similar in magnitude to what is obtained 
with an `effective coupling' \cite{DAS96}. For phenomenological 
analyses we suggest that the overall normalization be fitted to the data.

\subsection{Restoring gluons}

In the above analysis, all power corrections 
are obtained as convolution with the quark fragmentation function. 
But the original motivation was to compute fermion 
loops only to trace the dominating non-abelian (gluon) contribution 
through the dependence on $N_f$. Eventually, the secondary quark 
contribution should therefore be reinterpreted as a gluon fragmentation 
contribution. 

Such a reinterpretation faces ambiguities, and we would like to single 
out the gross features, which can be considered as unique. To have a 
basis for comparison, we first `deconvolute' the $g\to q\bar{q}$ 
splitting from the secondary quark contribution. We define 
the `effective gluon' coefficient functions 
$A^{g \leftarrow q}_{2,P}(x)$ through
\begin{equation}
\left[A^{g \leftarrow q}_{2,P}\ast P_{g\to q}\right](x) = 
A_{2,P}^{q,[s]}(x),
\end{equation}
which should be convoluted with the gluon fragmentation function, 
replacing $A_{2,P}^{q,[s]}$ convoluted with the quark fragmentation 
function. The superscript $g \leftarrow q$ indicates that the present
effective gluon coefficient has been obtained  by deconvoluting a quark
emission amplitude. Below we will consider an analogous coefficient obtained
from a scalar quark emission amplitude. Although
\begin{equation}
A^{g \leftarrow q}_{2,P}\ast D_g = A_{2,P}^{q,[s]}\ast D_q
\end{equation}
holds only to leading logarithmic accuracy in $Q^2$, this is the 
closest one can get to reinterpreting the secondary quark contribution 
as a gluon contribution. The functions $A^{g \leftarrow q}_{2,P}(x)$ 
can be compared directly with the $A_{2,P}^{g,\rm MG}(x)$ obtained 
in the massive gluon calculation.

To test the sensitivity of the deconvolution to the particular 
structure of the quark-gluon vertex, we compute $A_{n,P}$ in a (fictitious)  
theory with scalar particles rather than fermions (quarks). Then  
we deconvolute the gluon-scalar splitting function $P_{g\to sq}(x) = 
x (1-x)/2$ to obtain the effective gluon coefficient 
$A^{g \leftarrow sq}_{2,P}(x)$. The gluon interpretation of the 
secondary quark contribution is justifed only if this function, 
after convolution with the gluon fragmentation function, 
leads to results compatible with those obtained starting from quarks.

To obtain the functions $A_{n,P}$ for scalars, one replaces (neglecting 
terms that vanish when contracted with ${\cal M}_{\mu\nu}$) the trace
\begin{equation}
\label{ferm}
\frac{2}{\beta_0^f}\cdot
\frac{N_f}{2}\,{\rm Tr}[\gamma_\nu\!\not\!k_1\gamma_{\nu'}\!\not\!k_2] = 
6\pi\,\left(-8 k_{1\nu} k_{1\nu'}-2 k^2 g_{\nu\nu'}\right)
\end{equation}
in (\ref{xi-distributions}) by 
\begin{equation}
\frac{2}{\beta_0^{sf}}\cdot
\frac{N_s}{2}\,(k_1-k_2)_\nu (k_1-k_2)_{\nu'} = 
24\pi\cdot 4 k_{1\nu} k_{1\nu'}.
\end{equation}
When the phase-space over $k_1$ and $k_2$ is integrated unweighted, 
the integrals of the previous two equations coincide. Because of the 
projection ${\cal P}_{P;\mu\mu'}^{[s]}$ in (\ref{xi-distributions}), 
however, Eqs.~(\ref{a2ls})--(\ref{a4lts}) for the secondary 
quark contribution are replaced by  
\begin{eqnarray}
\label{sq1}
A_{2,L}^{sq,[s]}(x) &=& \frac{2}{5x^3}-\frac{2}{x}+2-\frac{2}{5}x^2~,\\
A_{4,L}^{sq,[s]}(x) &=& -\frac{6}{35x^5}+\frac{2}{5x^3}
-\frac{2}{5}+\frac{6}{35}x^2~,
\\
\label{sq2}
A_{2,L+T}^{sq,[s]}(x) &=& \frac{2}{5x^3}-\frac{2}{x}+2-\frac{2}{5}x^2~,
\\
A_{4,L+T}^{sq,[s]}(x) &=& -\frac{9}{35x^5}+\frac{4}{5x^3}-\frac{1}{x}
+\frac{2}{5}+\frac{2}{35}x^2~,
\end{eqnarray}
for scalar particles. Note that $A_{2,T}^{sq,[s]}(x)=0$. 

For the interesting case of a gluon splitting into two 
gluons, the expression replacing (\ref{ferm}) is 
still a linear combination of only two Lorentz structures 
$k_{1,\nu} k_{1,\nu'}$ and $k^2 g_{\nu\nu'}$. Therefore all ambiguities 
related to restoring $\beta_0$ are already exhausted by a linear 
combination of the results obtained for fermions and for scalars. In 
particular, if one computes the graphs with one cut gluon loop 
in an axial gauge, the coefficient of $k_{1\nu}k_{1\nu'}$ is independent 
of the gauge-fixing vector.

\subsection{Discussion: longitudinal fragmentation}

%%%%%%%%%%%%%%%%
% FIGURE 6
%%%%%%%%%%%%%%%%
\begin{figure}[p]
   \vspace{0cm}
   \centerline{\epsffile{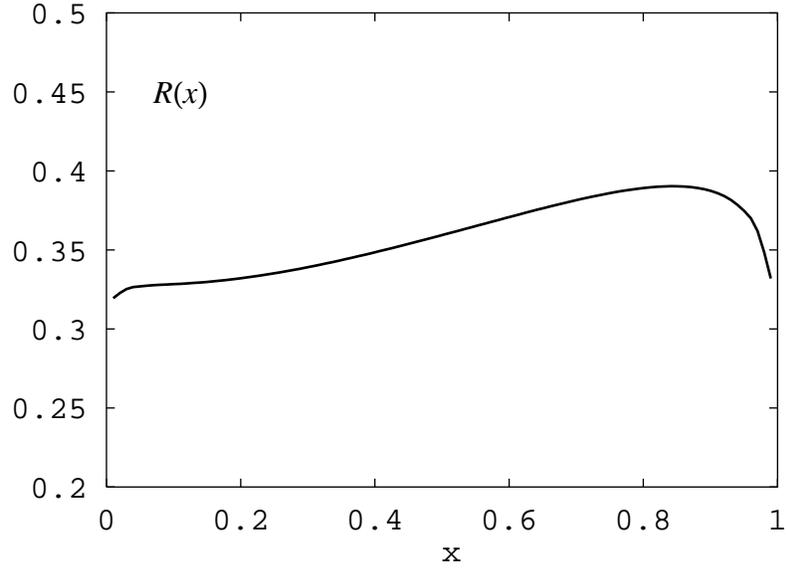}}
   \vspace*{0cm}
\caption{\label{ratioLong}
Ratio of secondary quark contribution in the fermion-loop 
approximation and gluon contribution in the massive gluon scheme.}
\end{figure}
%%%%%%%%%%%%%%%%
% FIGURE 7
%%%%%%%%%%%%%%%%
\begin{figure}[p]
%   \vspace{-3cm}
%   \epsfysize=6cm
%   \epsfxsize=10cm
   \centerline{\epsffile{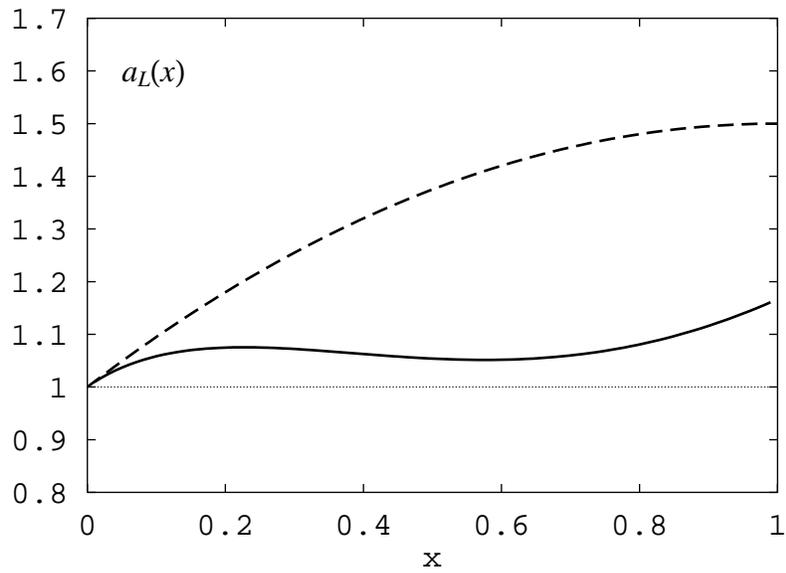}}
   \vspace*{0cm}
\caption{\label{deconvLong}
$a_L^q(x)$ (solid), $a_L^{sq}(x)$ (dashed) as defined in the text, compared 
with the massive gluon scheme (dotted).}
\end{figure}

We now consider the secondary quark contribution to longitudinal 
fragmentation in more detail. First, we compare the secondary quark 
contribution (see Fig.~\ref{figLong}) 
with the corresponding contribution in the 
massive gluon scheme. The ratio
\begin{equation}
R(x)=\frac{A_{2,L}^{q,[s]}\ast D_q(x)}{A_{2,L}^{g,\rm MG}\ast D_g(x)} 
\end{equation}
is plotted in Fig.~\ref{ratioLong}. Note that the fermion-loop approximation 
and massive gluon scheme coincide for the primary quark contribution, 
while  the secondary quark contribution is smaller by a factor of about   
3 than its massive  gluon analogue.
The difference in absolute normalization 
would be meaningful only  if we tried to fix the constants $K_2^i$ in terms 
of a single (universal) number as in \cite{DAS96}. 
If the $K_2^i$ are considered as adjustable parameters as proposed here, 
the important message conveyed by Fig.~\ref{ratioLong} is 
that the ratio $R$ is flat, so that the same $x$ dependence is obtained 
with the two methods. The remaining variation is  insignificant 
within the experimental uncertainties in the parametrization of 
leading-twist fragmentation functions and the theoretical uncertainties 
of our method. 

Second, we compare the effective gluon coefficients obtained 
from secondary quarks and scalars as described above. The coefficients 
are given by\footnote{To compare absolute normalizations, the effective 
gluon coefficient obtained from quarks has to be divided by 3 and that 
obtained from scalars by 12 to take care of additional factors 
as in (\ref{forty}).}
\begin{eqnarray}
A^{g \leftarrow q}_{2,L}(x) &=&
\frac{72}{7x^3}+18\frac{\ln x}{x}-\frac{15}{2x}
-\frac{39}{14}\sqrt{x}\cos\Big[\frac{\sqrt{7}}{2}\ln x\Big]~,
\nonumber\\
&&{}-\frac{3\sqrt{7}}{2}\sqrt{x}\sin\Big[\frac{\sqrt{7}}{2}\ln x\Big]
\\
A^{g \leftarrow sq}_{2,L}(x) &=&
    \frac{8(1-x)}{x^3}\left(2+ 2 x - x^2\right)~,
\end{eqnarray}
for quarks and scalars, respectively. We recall that the gluon coefficient 
in the massive gluon scheme is simply $8 (1-x)/x^3$, see (\ref{a2gmg}). 
To compare the $x$ dependence on a reasonable scale, we define
\begin{eqnarray}
A^{g \leftarrow q}_{2,L} (x) &=& \frac{8 (1-x)}{x^3}\cdot
\frac{7}{8}\cdot a_L^q(x)~,
\\
A^{g \leftarrow sq}_{2,L} (x) &=& 
\frac{8 (1-x)}{x^3}\cdot
\frac{1}{2}\cdot a_L^{sq}(x)~.
\end{eqnarray}
The normalizing factors are chosen such that $a_L(0)=1$. 
The $x$ dependence of $a_L$ is shown in Fig.~\ref{deconvLong} for all 
three procedures (massive gluon, fermion or scalar loops). We observe 
again very little $x$ dependence of the residual functions $a_L(x)$ and 
conclude that the method yields a unique result 
for $1/Q^2$ power corrections to longitudinal fragmentation. The 
residual different $x$ dependence is definitely beyond any accuracy that can 
be expected from the model. 

\subsection{Discussion: transverse fragmentation}

For the transverse fragmentation function, the effective gluon 
coefficients read
\begin{eqnarray}
A^{g \leftarrow q}_{2,T} (x) &=&
-18\frac{\ln x}{x}-\frac{51}{2x} +12
+\frac{15}{2}\sqrt{x}\cos\Big[\frac{\sqrt{7}}{2}\ln x\Big]~,
\nonumber\\
&&{}+\frac{27}{2\sqrt{7}}\sqrt{x}\sin\Big[\frac{\sqrt{7}}{2}\ln x\Big]
\\
A^{g \leftarrow sq}_{2,T} (x) &=& 0.
\end{eqnarray}
We have written the result for the transverse part separately, rather 
than the sum of transverse and longitudinal (referred to as `total' in 
this subsection), in order to emphasize the absence of a $1/x^3$ term. 
Being absent for quarks and scalars, it follows that it is absent 
for gluon loops as well in any gauge; therefore, the statement is 
independent of how precisely the non-abelian contribution is 
reinstated. 
On the other hand, cancellation of the leading $1/x^3$ terms is most likely
specific for the approximation of considering one chain of fermion bubbles
and, in this sense, accidental as a pecularity of the box graph.
For $1/Q^4$ 
corrections, the leading term for small $x$, which is $1/x^5$ in this 
case, is indeed present for both transverse and longitudinal coefficients, 
see $A_{4,P}^{q,[s]}$ in Sect.~2.4.2.
Since in large orders 
of perturbation theory diagrams with two chains of fermion loops 
are not suppressed, the cancellation that we observe does not imply,
strictly speaking, any parametric suppression 
of the leading power correction for transverse fragmentation.
A numerical suppression can hold, however, and it would be interesting
to see whether it can be inferred from experimental data.
Such a numerical suppression would be natural for the 
effective coupling approach of \cite{DOK95,DMW}.  

Comparing these results with the massive gluon scheme, we observe that 
in this scheme the leading term $1/x^3$ at small $x$ is cancelled in 
the sum of longitudinal and transverse coefficients, see (\ref{a2ltmg}). 
This cancellation appears to be unphysical and related to the fact 
that the structure of the massive gluon fragmentation cross section 
(\ref{ltmg}) is over-simplified. Indeed, the analytic terms in the 
$\xi$-expansion of (\ref{ltmg}) all behave as $\xi^n/x^{2n+1}$. On the 
other hand, non-analytic terms can arise only from the coefficient 
of the $\ln 1/(4\xi)$-term, which does not have a non-trivial 
expansion in $\xi$. All terms $\xi^n\ln\xi$ vanish for 
$n>3$ in this scheme, another indication that the massive gluon scheme 
may not capture the generic $x$ dependence of power corrections in the case 
of total fragmentation functions.\footnote{We discuss in Sect.~4 that 
the presence of $1/(Q^2 x^2)^n$ power corrections for the total 
fragmentation function does not conflict with the absence of $1/Q$ power 
corrections to the integrated gluon fragmentation cross section.}

%%%%%%%%%%%%%%%%
% FIGURE 8
%%%%%%%%%%%%%%%%
\begin{figure}[t]
   \vspace{0cm}
   \centerline{\epsffile{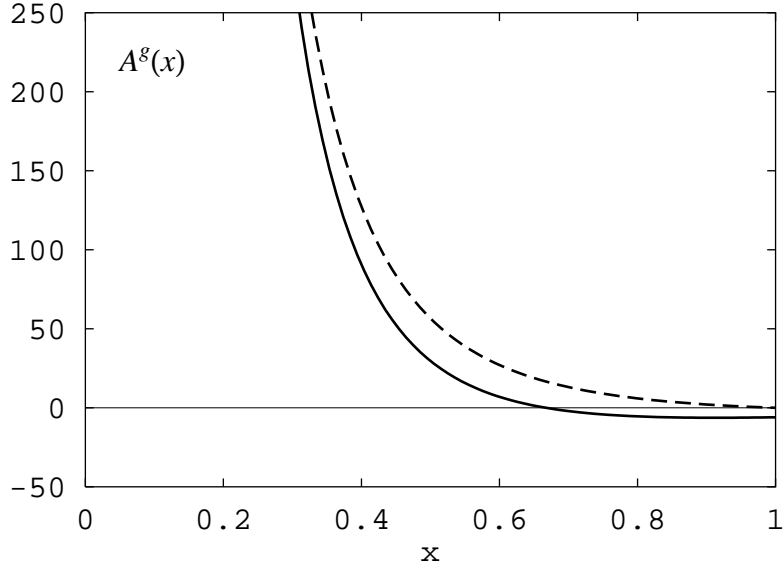}}
   \vspace*{0cm}
\caption{\label{deconvTotal}
Comparison of effective gluon coefficients obtained from quarks (solid) 
and scalars (dashed) for the sum of longitudinal and transverse 
fragmentation.}
\end{figure}

We compare the $x$ dependence of the effective gluon coefficients obtained 
from quark and scalar loops for the total fragmentation cross section 
in Fig.~\ref{deconvTotal}. The relative normalization is chosen 
such that the asymptotic behaviours at small $x$ coincide. 
Compared to Fig.~\ref{deconvLong}, we 
have not scaled out the small-$x$ behaviour, because we want to 
emphasize that the qualitative differences at large $x$ will be 
insignificant for practical applications after convolution with the 
gluon fragmentation function, because primary quark fragmentation 
dominates in this region. Opposite to longitudinal fragmentation, 
the large-$x$ behaviour of 
tranverse secondary quark fragmentation 
is not unambiguously predicted by our model. 
While the scalar result vanishes linearly at $x=1$, the quark result 
approaches a constant. 

Summarizing this discussion, we recommend the usage of the same functional 
form for the gluon coefficient in total fragmentation and in longitudinal 
fragmentation, but with a separate adjustable multiplicative constant
(recall that for scalar quarks, in the present approximation, the
two coefficients are in fact equal [Eqs.~(\ref{sq1}) and (\ref{sq2})], 
while for fermionic quarks they differ only slightly [Eqs.~(\ref{a2ls}) and
(\ref{a2lts})]). If the adjustable constant for total fragmentation turns 
out to be close to the one for longitudinal fragmentation, it would 
indicate that the cancellation of $1/x^3$ terms 
for transverse fragmentation observed in our approximation is 
operative to a degree. If it turned out much smaller than for 
longitudinal fragmentation, this would provide support for the 
cancellation observed in the massive gluon scheme. In addition, we 
may add a constant to account for the $x\to 1$ behaviour. However, 
the value of this constant is not important in practice, unless it 
were abnormally large. 

\subsection{A simple parametrization}

We summarize the analysis of this section in the form of a 
simple parametrization of $1/Q^2$ power corrections, which we 
suggest to apply to the analysis of fragmentation data. 
We write the fragmentation cross section 
(\ref{withpower}) as 
\begin{equation}
\frac{d\sigma_P}{dx}(x,Q^2) = \frac{d\sigma_P^{\rm LT}}{dx}(x,Q^2) + 
\frac{d\sigma_P^{\rm power}}{dx}(x,Q^2).
\end{equation}
The leading-twist cross section is as in (\ref{DGLAP}). The $1/Q^2$ power 
corrections are parametrized as
\begin{eqnarray}
\frac{d\sigma_L^{\rm power}}{dx}(x,Q^2) &=& 
\frac{1 \mbox{GeV}^2}{Q^2}\int_x^1\frac{dz}{z}\,
\Bigg\{c_{q,L}\left[\delta(1-z)+\frac{2}{z}
\right]\,D_q(x/z,\mu) \nonumber\\
&&\,+ c_{g,L}\,\frac{1-z}{z^3}\,D_g(x/z,\mu)
\Bigg\}~,
\label{parL}
\\
\frac{d\sigma_{L+T}^{\rm power}}{dx}(x,Q^2) &=& 
\frac{1 \mbox{GeV}^2}{Q^2}\int_x^1\frac{dz}{z}\,
\Bigg\{c_{q,L+T}\left[-\frac{2}{[1-z]_+} + 1 + 
\frac{1}{2}\delta'(1-z)\right]\,D_q(x/z,\mu) 
\nonumber\\
&&\,+ \left[c_{g,L+T}\,\frac{1-z}{z^3} + d\right]\,D_g(x/z,\mu)
\Bigg\}~,
\label{parLT}
\end{eqnarray} 
where $D_{i}$ denotes the leading-twist fragmentation function for parton 
$i$ to decay into any given hadron, and the plus distribution is defined as 
usual,
\begin{equation}
\int\limits_0^1 dx\,\frac{f(x)}{[1-x]_+} = 
\int\limits_0^1 dx\,\frac{f(x)-f(1)}{1-x}~,
\end{equation}
for any test function $f$. The coefficients in front of quark fragmentation 
functions are taken from Sect.~2.4.1. The coefficients in front of 
gluon fragmentation functions follow from the discussion in Sects.~3.3 and 
3.4. Note that for longitudinal fragmentation, we have neglected the 
residual $x$ dependence displayed in Fig.~\ref{deconvLong} and chosen the 
simplest functional form. Let us add the following remarks:

(a) The overall scale $1\,\mbox{GeV}/Q$ ($Q$ in GeV) has been chosen 
to set the overall magnitude for the constants $c_k$ and $d$. With this 
scale, we expect these constants to be of order 1.

(b) The constants $c_k$ and $d$ should be fitted from the data. As 
mentioned above, the constant $d$ is actually insignificant, so that 
the parametrization depends on four constants only. It is 
important that these constants should be quoted only in connection 
with a value for $\mu$ chosen for the factorization scale in 
the leading-twist contribution and also together with the order of 
perturbation theory to which the leading-twist coefficients 
$C_P^i$ have been used. In this sense, the constants should be 
considered $\mu$-dependent. We recall that the fermion-loop approximation
considered in this paper suggests $c_{g,L}\simeq c_{g,L+T}$ while
in the gluon mass scheme $c_{g,L} \gg c_{g,L+T}$.

(c) Since we did not consider additional logarithmic $Q$ dependence 
of power corrections, the scale $\mu$ in $D_i$ above is undetermined. 
Without additional information, it is  natural to set 
$\mu$ equal to the scale used in the leading-twist part of the 
cross section.

(d) If the above parametrization of power corrections is used, no 
Monte Carlo correction for hadronization should be applied in addition.

(e) The ansatz can only be used as long as $x>\Lambda/Q$, where 
$\Lambda\sim 1\,$GeV is a typical QCD scale. At smaller values 
of $x$, higher power corrections such as $1/Q^4$ become as important 
as $1/Q^2$ corrections. In fact, the $1/Q^2$ expansion breaks down 
and would have to be resummed. In other words, (\ref{parL}) and 
(\ref{parLT}) are corrections to the leading-twist prediction. If 
the power correction becomes as large as the leading-twist result, 
the ansatz fails.

%%%%%%%%%%%%%%%%%%%%%
% SECTION 4
%%%%%%%%%%%%%%%%%%%%%

\section{$1/Q$ corrections to integrated cross sections}

For $x>\Lambda/Q$ and not too close to 1, the power corrections 
discussed in the previous section arise only from regions of integration 
where two partons are collinear. In this section we consider moments 
of the $x$ distributions, specifically the second moment (\ref{moment2}), 
related to the integrated longitudinal and transverse 
cross sections, which are infrared finite. 

The moments can have qualitatively different power corrections from those 
at finite $x$, because they include new infrared sensitive regions 
$x\to 0$ and $x\to 1$, related to soft partons. For small $x$ this 
possibility is evident from the increasingly divergent terms 
$1/(Q^2 x^2)^n$ found in the secondary quark or gluon contributions.  
Thus, the integration over $x$ must be performed before expansion 
in $\xi$. In other words, an infinite series of power corrections 
in $1/Q^{2n}$ has to be resummed to obtain the integrated cross 
section. Specifically, for the gluon contribution in the massive 
gluon scheme
\begin{equation}
\label{Q}
\int\limits_{\sqrt{\xi}} dx\,\frac{1}{2}x\,\frac{\xi^n}{x^{2 n+1}} 
\sim \sqrt{\xi}
\end{equation}
for any $n$. Recalling that $\xi$ should be interpreted as $\Lambda^2/Q^2$, 
we see that the small-$x$ region can produce a $1/Q$ power correction, 
parametrically larger than at finite $x$.

In Sect.~4.1 we collect the distributions in $\xi$ for the various 
contributions to the integrated cross section. Sect.~4.2 contains some 
observations on the relation of small-$x$ behaviour and 
$1/Q$ corrections.

\subsection{Longitudinal and transverse cross section}

The primary quark contribution to the longitudinal and transverse 
cross sections is given by
\begin{eqnarray}
 \frac{\sigma_L^{q,[p]}}{\sigma_0}
 &=&\frac{C_F\alpha_s}{2\pi}\cdot 2\cdot\frac{1}{4}
\Bigg[(1-\xi)(1+31\xi-2\xi^2)+12\xi\ln\xi+6\xi^2(3\ln\xi-\ln^2\xi)\Bigg]
\nonumber\\
&=& \frac{C_F\alpha_s}{2\pi}\cdot 2\cdot
\Bigg[\frac{1}{4}+\xi\Big(3\ln\xi+\frac{15}{2}\Big)
+\xi^2\Big(-\frac{3}{2}\ln^2\xi +\frac{9}{2} \ln\xi-\frac{33}{4}\Big)
+\ldots\Bigg]~,
\end{eqnarray}
\begin{eqnarray}
\frac{ \sigma_T^{q,[p]}}{\sigma_0}
 &=& \frac{C_F\alpha_s}{2\pi}\cdot 2\cdot
\Bigg\{\frac{2}{3}\ln\xi +\frac{22}{9}-7\xi+\frac{17}{4}\xi^2-\frac{22}{9}\xi^3
+\xi\ln\xi -\frac{9}{2}\xi^2\ln\xi
\nonumber\\
&&{}+\frac{2}{3}\xi^3\ln\xi-(1+\xi)^2\Bigg[\ln\xi\ln(1+\xi)
-\frac{1}{2}\ln^2(1+\xi)-{\rm Li}_2\Big(\frac{\xi}{1+\xi}\Big)\Bigg]\Bigg\}
\nonumber\\
&=& \frac{C_F\alpha_s}{2\pi}\cdot 2\cdot
\Bigg[\frac{2}{3}\ln\xi +\frac{22}{9}-6\xi
+\xi^2\Big(\frac{3}{2}\ln^2\xi -\frac{3}{2} \ln\xi+6\Big)
+\ldots\Bigg]~.
\end{eqnarray}
The primary quark contribution does not contain odd powers of 
$\sqrt{\xi}$ and therefore no $1/Q$ power correction.

Because of large cancellations, it is again useful to present the 
distributions for the longitudinal and total secondary quark contribution. 
We have 
\begin{eqnarray}
\frac{\sigma_L^{q,[s]}}{\sigma_0}
 & = & \frac{(1 - \xi)(2 - 13 \xi + 
2 \xi^2)}{4}
 -  \frac{9 \xi (1 + \xi)}{8} \ln \xi 
-\frac{15 \xi (1 - \xi)}{32} \ln^2 \xi 
\nonumber\\
&&{}+ \left[ \frac{15}{16} (1 + \xi^2) + \frac{27}{8} \xi 
\right]\sqrt{\xi} S(\sqrt{\xi}) 
\nonumber\\
&=&\frac{C_F\alpha_s}{2\pi}\Bigg[
1-\frac{15\pi^3}{64}\sqrt{\xi}-6\xi\ln\xi-\frac{27\pi^3}{32}\xi^{3/2}-
\frac{46}{3}\xi^2\ln\xi 
\nonumber\\
&&{}+\frac{308}{9}\xi^2-\frac{15\pi^3}{64}\xi^{5/2}+\ldots\Bigg]~,
\label{lorenzo} 
\end{eqnarray}
where
\begin{eqnarray}
S(\lambda) & = & - \frac{\pi^3}{8} 
+ 2 \lambda \int_0^1\frac{\ln^2(\lambda t)\,dt}{1+\lambda^2 t^2}
= - \frac{\pi^3}{8} + 
{\rm i} \ln^2\lambda \ln \frac{1-{\rm i}\lambda}{1+{\rm i}\lambda}
%2 \arctan \lambda \, \ln^2 \lambda 
\nonumber \\
& + & 2 {\rm i} \left[\ln \lambda \left( {\rm Li}_2 ({\rm i} \lambda) -
{\rm Li}_2 (- {\rm i} \lambda) \right) - 
\left( {\rm Li}_3 ({\rm i} \lambda) - {\rm Li}_3 (- {\rm i} \lambda) \right)
\right]~,
\label{lis}
\end{eqnarray}
and no further half-integer powers of $\xi$ appear in (\ref{lorenzo})
beyond $\xi^{5/2}$.
The total secondary quark contribution is simply
\begin{equation}
\frac{\sigma_{L+T}^{q,[s]}}{\sigma_0} = \frac{C_F\alpha_s}{2\pi}\Bigg[
-\frac{4}{3}\ln\xi -\frac{35}{9}-6\xi\ln\xi-3\xi-6\xi^2\ln\xi
+3\xi^2-\frac{4}{3}\xi^3\ln\xi+\frac{35}{9}\xi^3\Bigg]~.
\label{totalg}
\end{equation}
In the massive gluon scheme, the gluon contribution, to be compared 
with the secondary quark contribution, reads
\begin{eqnarray}
\frac{\sigma_L^{g,{\rm GM}}}{\sigma_0}
&=&  \frac{C_F\alpha_s}{2\pi}
(1+\xi)\Bigg\{1-\xi^2+2\xi\ln\xi+\xi\ln^2\xi
\nonumber\\
&-&4\sqrt{\xi}\Bigg[(1+\sqrt{\xi})^2
\Big(\frac{1}{2}\ln\xi\ln(1+\sqrt{\xi})
+\frac{\pi^2}{12}+{\rm Li}_2(-\sqrt{\xi})\Big)
\nonumber\\
&&{}+(1-\sqrt{\xi})^2
\Big(-\frac{1}{2}\ln\xi\ln(1-\sqrt{\xi})
+\frac{\pi^2}{6}+{\rm Li}_2(\sqrt{\xi})\Big)\Bigg]\Bigg\}
\nonumber\\
&=&  \frac{C_F\alpha_s}{2\pi}
\Bigg[1-\pi^2\sqrt{\xi}+\xi\Big(\ln^2\xi-2\ln \xi+9+\frac{2\pi^2}{3}\Big)
-2\pi^2\xi^{3/2}
\nonumber\\
&&{}+\xi^2\Big(\ln^2\xi-\frac{10}{3}\ln\xi +\frac{107}{9}+\frac{2\pi^2}{3}\Big)
+\ldots\Bigg]~,
\\
\sigma_{L+T}^{g,{\rm GM}} &=& \sigma_{L+T}^{q,[s]}~~.
\end{eqnarray} 
Both the secondary quark contribution and the gluon contribution in the 
massive gluon scheme contain a $\sqrt{\xi}$ term, although with different 
coefficient. This term cancels in the sum of the longitudinal and transverse 
cross sections. Summing $\sigma_{tot} = \sigma_L^{q,[p]}+\sigma_T^{q,[p]}+
\sigma_{L+T}^{q,[s]}$ we reproduce the distribution function for the 
total cross section given in Appendix~B of Ref.~\cite{BEN95}. 
Note that because the longitudinal and transverse cross sections are 
infrared finite, we continued to neglect explicit infrared factorization. 
However, the secondary quark contribution now contains the order-$\alpha_s$ 
gluon contribution.

\subsection{Discussion}

We now discuss some general aspects of how $1/Q$ corrections are generated 
in the gluon (secondary quark) contribution to the integrated cross section. 
As far as these general aspects are concerned, the fermion-loop approximation 
and the massive gluon scheme are the same and we choose the latter for 
the discussion, because the analytic expressions are simpler in this 
case.

We noted above, in Eq.~(\ref{Q}), that $1/Q$ corrections are made possible, 
because the expansion parameter at finite but small $x$ turned out to 
be $\xi/x^2$ rather than $\xi$. To obtain the coefficient of 
$\sqrt{\xi}$ in the integrated cross section, we need to pick up the 
most singular term in $1/x$ at every order in $\xi$. Once this is 
realized, the real question is not why $1/Q$ corrections arise in the 
integrated cross section, but why they do not arise in the sum of the
longitudinal and transverse cross sections. Indeed, the most singular 
terms in the expansion of the total gluon cross section (\ref{ltmg}) 
are
\begin{equation}
\frac{1}{\sigma_0}\frac{d\sigma_{L+T}^{g,{\rm MG}}}{dx} =
\frac{C_F\alpha_s}{2\pi}\,\frac{4}{x}\left[\ln\frac{x^2}{\xi}\,-\,
\sum_{n=1}^\infty\frac{(2 n)!}{(n!)^2 n}\left(\frac{\xi}{x^2}\right)^n\,
\right],
\end{equation}
and the general structure of the expansion is the same as for the 
longitudinal and transverse cross sections separately. Integrating this 
expansion term by term, we obtain, for the coefficient $a_1$ of 
$\sqrt{\xi}$ in the expansion of $\sigma_{L+T}^{g,{\rm MG}}$, the
expression
\begin{equation}
a_1=4\left[2 (1-\ln 2)\,-\,\sum_{n=1}^\infty\frac{(2 n)!}{(n!)^2 n}
\frac{2^{-2 n}}{2 n-1}\right].
\end{equation}
The sum is evaluated to 
\begin{eqnarray}
&&\sum_{n=1}^\infty\frac{(2 n)!}{(n!)^2 n}\frac{2^{-2 n}}{2 n-1} = 
-\sum_{n=1}^\infty\frac{1}{n! n}\frac{\Gamma(n-1/2)}{\Gamma(-1/2)} 
\nonumber\\
&&=\,\lim_{\epsilon\to 0}\frac{1}{\epsilon}\left[1-{}_2 F_1(-1/2,\epsilon,
1+\epsilon;1)\right] = 2 (1-\ln 2),
\end{eqnarray}
so that $a_1=0$, as is evident from (\ref{totalg}). From this exercise we 
see that the cancellation of $1/Q$ corrections in the total gluon 
contribution is a property of the power expansion at fixed $x$ only when 
all orders are included. Had we truncated the sum over $n$, a non-zero result 
would have been obtained. Furthermore, it is necessary to consider also 
analytic terms in the $\xi$ expansion at fixed $x$ to obtain the 
cancellation. This means that while $1/Q$ corrections are tied to 
soft partons, collinearity is not important. 

We would like to note that these properties are reminiscent of those 
that lead to a cancellation of $1/Q$ corrections in the Drell-Yan 
cross section in the same (fermion-loop) approximation \cite{BBDY}. 
Contrary to the Drell-Yan process, however, we know for fragmentation 
that $1/Q$ corrections must cancel after summing over contributions 
from all partons, because the total $e^+ e^-$ annihilation cross 
section receives at most $1/Q^4$ power corrections (leaving aside 
issues related to parton-hadron duality). On the other hand, the 
cancellation of $1/Q$ terms in $\sigma^g_L+\sigma^g_T$ and $1/Q^2$ terms 
in $\sigma^g_{L+T}+\sigma^q_{L+T}$ looks rather accidental from the 
diagrammatic point of view.

The fact that the power expansion at fixed but small $x$ runs in 
$\xi/x^2$ reminds us that the scale $\Lambda$ of non-perturbative effects 
must be compared not with the total energy $Q$ of the process, but with 
the energy $(Q x)/2$ of the detected hadron. To see this more 
transparently, consider the longitudinal gluon contribution 
(\ref{lmg}). We may simplify the factor $(1+\xi) (1+\xi-x)$ to 1, because 
it does not alter $1/Q$ corrections, and we omit the factor $C_F\alpha_s/
(2 \pi)$. Now $x$ times (\ref{lmg}) depends on $\xi$ and $x$ only in 
the combination $x/\sqrt{\xi}$. The Borel transform $B[F]$ of the 
perturbative expansion generated by cut and virtual fermion-loop 
diagrams is related to the $\xi$ distribution $F$ by the 
Mellin transform \cite{BEN95}:
\begin{equation}
B[F](u,x) = -\frac{\sin\pi u}{\pi u}\int\limits_0^1 
d\xi\,\xi^{-u}\,\frac{d}{d\xi}F(\xi,x).
\label{mellin}
\end{equation}
Because of the dependence on $x/\sqrt{\xi}$ only, the $x$ dependence 
factorizes in the Borel transform as
\begin{equation}
B\left[\frac{1}{2}x\frac{d\sigma^{g,{\rm MG}}}{\sigma_0 dx}
\right] = \left(\frac{x}{2}\right)^{-2 u}\,B(u),
\label{bor}
\end{equation}
with an $x$-independent function $B(u)$, which is easily obtained 
from (\ref{lmg}) and (\ref{mellin}). Its precise form will not be needed, 
but we note that $B(u)$ is analytic for $|u|<1$ with a pole at $u=1$ 
related to $1/Q^2$ power corrections.   
The (formal) Borel representation is now given by
\begin{eqnarray}
\frac{1}{2}x\frac{d\sigma^{g,{\rm MG}}}{\sigma_0 dx} &=& 
\left(-\frac{1}{\beta_0}\right)\,
\int\limits_0^\infty d u\,\exp\left(-\frac{u}{(-\beta_s\alpha_s(Q))}
\right)\,\left(\frac{x}{2}\right)^{-2 u}\,B(u)
\nonumber\\
&=&\left(-\frac{1}{\beta_0}\right)\,
\int\limits_0^\infty d u\,\exp\left(-\frac{u}{(-\beta_s\alpha_s(Qx/2))}
\right)\,B(u).
\end{eqnarray}
The $x$ dependence is completely accounted for by 
setting the scale of the coupling constant to $Q x/2$, the energy 
of detected particles in the final state. Moreover, if we adhere to 
a strictly perturbative approach, the Borel integral diverges at 
infinity for $x<2\Lambda/Q$ due to the Landau pole in the (one-loop) 
running coupling. Sensitivity to $u=\infty$ indicates that 
the entire power expansion in $1/Q^2$ needs to be resummed and that 
the renormalon approach is inapplicable at such small $x$.

Let us now define moments in $x$ with a cut on $x$ that eliminates the 
small-$x$ region:
\begin{equation}
\sigma^{g,{\rm MG}}(\gamma,x_c)\equiv\frac{1}{\sigma_0}
\int\limits_{x_c}^1 dx\,\frac{1}{2}\,x^{1+\gamma}\,
\frac{d\sigma^{g,{\rm MG}}}{dx}.
\end{equation}
Eq.~(\ref{bor}) gives
\begin{equation}
B[\sigma^{g,{\rm MG}}(\gamma,x_c)](u) = 
\frac{1-x_c^{1+\gamma-2 u}}{1+\gamma-2 u}\,2\,B(u).
\label{xc}
\end{equation}
Taking first $x_c=\gamma=0$, we see that the $1/Q$ correction 
to the integrated longitudinal gluon cross section (corresponding to a 
pole of the Borel transform at $u=1/2$) again is an immediate consequence of 
the scale being $Q x$. If the scale were $Q\sqrt{x}$, the denominator 
would read $1/(1-u)$ instead of $1/(1-2 u)$.

Keeping $\gamma=0$, we note that the pole at $u=1/2$ is eliminated 
no matter how small $x_c$, as long as it is non-zero. Since in any 
experimental situation a minimum energy cut is required, the presence 
of $1/Q$ corrections in measured quantities depends on the value of 
$x_c$. If $x_c\sim\Lambda/Q$, the Borel transform is sharply 
peaked at $u=1/2$ and the structure of large-order perturbation theory 
is indistinguishable from a $1/Q$ IR renormalon for practical 
purposes.

Taking $x_c=0$, we note that the position of the pole depends on 
the moment. In particular, depending on $\gamma$, it is not bound 
to be integer or half-integer, a situation that never occurs 
in deep-inelastic scattering. It is this dependence on the precise 
weighting of long-distance regions (small $x$), first noted in 
\cite{MW}, that makes $1/Q$ power corrections to $\sigma_L$ and event 
shapes in general difficult to interpret in terms of operators 
(as in DIS) in the conventional sense. It also prompts some 
caution regarding the final conclusion on $1/Q$ corrections. It is 
conceivable that resummation of logarithmic terms $\ln^n x$ in 
perturbation theory modifies the approach to the small-$x$ region. 
If, for illustration, resummation would turn $x^{-2 u}$ into 
$x^{\gamma-2 u}$ in (\ref{bor}), a fractional power-correction 
would accordingly be obtained from (\ref{xc}). However, at present 
we do not know how to combine small-$x$ resummation and 
constraints from angular ordering with a 
renormalon analysis. 

The resummation of leading $\ln x$ can formally be achieved by a 
modification of the DGLAP evolution equation, which 
replaces the fragmentation 
functions $D_i(x/z,Q^2)$ with $D_i(x/z,x^2 Q^2)$ under the convolution 
integral~\cite{WLEC}. The evolution equation then reads
\begin{equation}
Q^2 \frac{\partial}{\partial Q^2} D_g(N,Q^2)=
\frac{\alpha_s(Q)}{2\pi}\int_0^1 dz\,z^{N-1} P_{g\to g}(z)D_g(N,z^2Q^2)~.
\end{equation} 
Trading $\alpha_s(Q)$ for $1/(-\beta_0 \ln Q^2)$ and 
writing the fragmentation function formally as a Mellin integral 
\begin{equation}
 D_g(N,Q^2) = \int_0^\infty du\, e^{-u/(-\beta_0\alpha_s(Q))} 
\tilde D(N,u)~,
\end{equation}
one can get the solution
\begin{equation}
 \tilde D(N,u)\propto u^{-1-2\gamma_{gg}/b}(2u+1-N)^{2\gamma_{gg}/b}
(2u-N)^{-4N_c/(Nb)}e^{1/b\, G(N,u)}~,
\label{suspect}
\end{equation}
where $\gamma_{gg} = 2 N_c/(N-1)$, $b=-4\pi\beta_0$ and $G(N,u)$ is analytic
at $N+1-2u>0$. We 
approximated the gluon-gluon splitting function by its dominant term at 
small $x$. As $N\to 1$, the singularities at $u=0$ and $u=(N-1)/2$ 
merge, and produce $\tilde D_g(N,u)\propto(1/u)\exp[-2N_c/(bu)]$,
which gives rise to the correct small-$x$ behaviour. 
On the other hand, (\ref{suspect}) exhibits singularities at $u=(N-1+k)/2$,
with $k=1,2,\ldots$, which might indicate non-perturbative effects through  
power-suppressed ambiguities. 
Such an interpretation is, however, speculative, and we will not pursue it
further in this paper. In general, one may ask, as for 
Drell-Yan production \cite{KS95,BBDY}, whether evolution equations 
of this type also give information on power corrections. In the above  
example, keeping only leading-$\ln x$ terms in 
each order of perturbation theory gives $2 B(0)$ as residue of 
the pole at $u=1/2$, while the correct result is $2 B(1/2)$, 
see (\ref{xc}). In general, one has to be aware that systematic 
procedures to sum $\ln x$ terms 
may not work to power accuracy.

%%%%%%%%%%%%%%%%%%%%%%%%
% SECTION 5
%%%%%%%%%%%%%%%%%%%%%%%%

\section{Perturbative corrections to $\sigma_L$ and the 
determination of $\alpha_s$}

A measurement of the integrated longitudinal cross section could eventually 
yield a precise determination of the strong coupling. 
To approach this goal theoretically one has to control higher-order 
perturbative corrections, as well as non-perturbative effects, both of which 
are expected to be much larger for $\sigma_L$ than for the total cross section 
$\sigma_{tot}$. 
In this section we consider perturbative corrections to $\sigma_L$, 
written as
\begin{equation}
\sigma_L = \sigma_0\,\frac{\alpha_s}{\pi}\,\left[1+
\sum_{n=0}^\infty d_n\,(-\beta_0\alpha_s)^n\right] ,
\end{equation}
with $\beta_0=-1/(4\pi)[11-2N_f/3]$ as before. 
We approximate the exact higher-order coefficient 
by its value in the `large-$\beta_0$' limit, where $\beta_0$ is restored 
from the term with the largest power of $N_f$ at each order. Given the 
$\xi$ distributions for the primary and secondary quark contributions to 
$\sigma_L$ in Sect.~4.1, the coefficients $d_n$ are obtained 
numerically from the logarithmic moment integrals given in 
Eq.~(\ref{logmoments}), as discussed in detail in \cite{BEN95}. 
The `large-$\beta_0$' approximation, called `naive non-abelianization' 
(NNA) in 
\cite{BEN95}, reduces to the BLM scale setting prescription \cite{BLM} for 
$n=1$. To see how it works, we rewrite the exact $\alpha_s^2$ correction in 
(\ref{sigmal}) as 
\begin{equation}
d_1=6.17-0.7573/(-\beta_0) .
\end{equation}
With $-\beta_0=0.61$ for $N_f=5$, neglecting the second term gives an 
accuracy of about $25\%$. In particular, the approximation predicted 
correctly the large size of the second-order correction. 

We have calculated the coefficients $d_n$ in higher
orders, in the $\overline{\mbox{MS}}$ scheme. The `primary' and 
`secondary' quark contributions, $d_n^{q,[p]}$ and $d_n^{q,[s]}$, 
respectively, add to $d_n$ as $d_n=d_n^{q,[p]}/3+2 d_n^{q,[s]}/3$. 
Columns 2 to 4 of Tab.~\ref{table} show the primary and secondary 
quark coefficients, together with successive finite-order 
approximations to $\sigma_L/\sigma_{tot}$ based on these coefficients. 
The perturbative coefficients  grow 
rapidly, especially for the secondary quark contribution. The fixed-sign 
growth and faster growth for the secondary quark contribution are directly 
related to an IR renormalon, indicating a $1/Q$ correction to 
secondary quark fragmentation as discussed in Sect.~4.

\begin{table}[t]
\addtolength{\arraycolsep}{0.3cm}
\renewcommand{\arraystretch}{1.3}
$$
\begin{array}{cccccc}
\hline\hline
 n & d_n^{q,[p]}\,(\overline{\mbox{MS}}) & 
     d_n^{q,[s]}\,(\overline{\mbox{MS}}) & 
     \sigma_{L,n}/\sigma_{tot}           &
     d_n^{q,[s]}\,(\mbox{V})      & 
     d_n^{sq,[s]}/d_n^{q,[s]}\,(\mbox{V}) \\ 
\hline
 0 & 1           & 1           & 0.036 & 1           & 1     \\
 1 & 11/2        & 13/2        & 0.052 & 29/6        & 1.121 \\
 2 & 29.82       & 45.97       & 0.060 & 27.07       & 1.203 \\ 
 3 & 164.1       & 369.0       & 0.064 & 188.8       & 1.226 \\
 4 & 944.1       & 3441        & 0.066 & 1634        & 1.219 \\
 5 & 5829        & 3.734\,10^4 & 0.068 & 1.703\,10^4 & 1.210 \\
 6 & 3.940\,10^4 & 4.682\,10^5 & 0.070 & 2.088\,10^5 & 1.205 \\
 7 & 2.948\,10^5 & 6.707\,10^6 & 0.071 & 2.954\,10^6 & 1.202 \\
 8 & 2.447\,10^6 & 1.086\,10^8 & 0.073 & 4.751\,10^7 & 1.201 \\\hline\hline
\end{array}
$$
\caption{\label{table}
Perturbative corrections to $\sigma_L$ as obtained from `naive 
non-abelianization'. The fourth column shows successive values for 
$\sigma_{L,n}/\sigma_{tot}$ at $Q=M_Z$ and with $\alpha_s(M_Z)=0.118$. 
The last two columns show a comparison of NNA coefficients 
in the $V$ scheme ($C=0$) obtained 
from fermions or scalars. The asymptotic ratio of scalar and 
fermion coefficients is $6/5$; $\sigma_{tot}/\sigma_0=1.04$ has been 
used.}
\end{table}

Since the longitudinal cross section is not fully inclusive
with respect to gluon splitting into a quark-antiquark pair, restoration
of the non-abelian piece is not unique, just as it was not unique 
for power corrections. The resulting ambiguity presents 
a major difficulty for the extension of `naive non-abelianization' 
(and BLM scale setting) to hadronic event shape observables. 
We have not found a physically motivated modification of NNA that
would alleviate this difficulty; therefore we investigated the ambiguity 
by comparing the higher-order coefficients obtained 
from restoring non-abelian contributions from both fermion and scalar 
loops.\footnote{This comparison was suggested to us by Lance Dixon.} 
The comparison
is shown in the last two columns of Tab.~\ref{table} in the V-scheme. 
To make a meaningful comparison in other schemes 
one would have to adjust the values of the strong coupling to take 
into account the difference
between the finite parts of the fermion (scalar) loop. (In the 
$\overline{\rm MS}$ scheme $C=-5/3$ for quarks and $C=-8/3$ for scalars.)     
It is seen that the ratio scalars/quarks is very stable around $1.2$, which 
is the asymptotic limit in large orders and corresponds to the ratio 
of the coefficients of $\sqrt{\xi}$ in the expansion of the 
secondary quark distribution function. Consequently, the ambiguity 
in restoring the non-abelian contributions is at least $20\%$. However, 
in higher orders, the NNA prescription is probably less accurate than this 
anyway. The 
main point here is that radiative corrections to $\sigma_L$ have the
same sign and are consistently large already in low orders. 
Keeping this in mind, we conform to 
restoring higher-order contributions from fermion loops in the following. 

The sum of $(\beta_0\alpha_s)^n$ contributions to all orders is
conveniently written in terms of `enhancement factors' \cite{BEN95}, 
measured relative to the leading-order contribution. They are   
defined by 
\begin{equation}
M^{[p,s]}(\alpha_s) = 1 + \sum_{n=0}^\infty (-\beta_0\alpha_s)^n 
d_n^{q,[p,s]}~,
\end{equation}
so that
\begin{equation}
 \sigma_L^{({\rm NNA})} =\sigma_0 \frac{\alpha_s}{\pi}
\left[\frac{1}{3}M^{[p]}+\frac{2}{3}M^{[s]}\right].
\end{equation}
In the following we define the sum to infinity in the sense of a 
principal value Borel integral, although truncating the series at 
its minimal term would be completely equivalent for practical purposes. 
For various values of $\alpha_s(M_Z)$ we get, at $Q=M_Z$,
\begin{eqnarray}\label{Mfactors}
&&\alpha_s =0.110: \qquad 
 M^{[p]}=1.59 \qquad M^{[s]} = 1.92\pm 0.05.
\nonumber\\  
&&\alpha_s =0.120: \qquad 
 M^{[p]}=1.68  \qquad M^{[s]} = 2.08\pm 0.08.
\\  
&&\alpha_s =0.130: \qquad 
 M^{[p]}=1.79  \qquad M^{[s]} = 2.23\pm 0.12.
\nonumber
\end{eqnarray}
The given uncertainties for the secondary quark 
contribution roughly coincide with the 
size of the minimal term in the series. The corresponding 
uncertainty for $M^{[p]}$ is small in comparison with the one for 
$M^{[s]}$ and is omitted. Interpolating between these values, one can 
get a rough idea of how  $\sigma_L^{({\rm NNA})}$ changes with 
$\alpha_s$. 

Experience with similar calculations suggests that the approximation 
of resumming only $(\beta_0\alpha_s)^n$ contributions 
overestimates radiative corrections. A more 
realistic estimate would be expected to lie 
in between the resummed and the exact
next-to-leading order (NLO) result. To see the magnitude of 
separate contributions, we take half 
of the Borel sum of  $(\beta_0\alpha_s)^n$ contributions starting at  
order $n=3$ to estimate the size of higher-order corrections.
With $\alpha_s(M_Z)=0.118$ we then get
\begin{equation}
\sigma_L/\sigma_{\rm tot} = 0.0495 + (0.010\pm 0.010) \pm 0.003~,
\end{equation}
where the first number corresponds to the NLO result, the second bracket 
presents our estimate
of further perturbative corrections, and the third gives the
estimated uncertainty in summation of the perturbation theory. 
At $Q=M_Z$ this uncertainty is rather moderate in size, even though 
it corresponds to a $1/Q$ correction. This suggests that although 
non-perturbative corrections to $\sigma_L$ are much larger than 
to $\sigma_{tot}$, these corrections are still 
not large at $Q=M_Z$. An exact $O(\alpha_s^3)$ calculation would reduce
the theoretical error considerably. Then, $\alpha_s$ determined 
from $\sigma_L$ and $\sigma_{tot}$ would provide 
an interesting consistency check.  

%%%%%%%%%%%%%%%%
% FIGURE 9
%%%%%%%%%%%%%%%%
\begin{figure}[t]
%   \vspace{-3cm}
%   \epsfysize=6cm
%   \epsfxsize=10cm
   \centerline{\epsffile{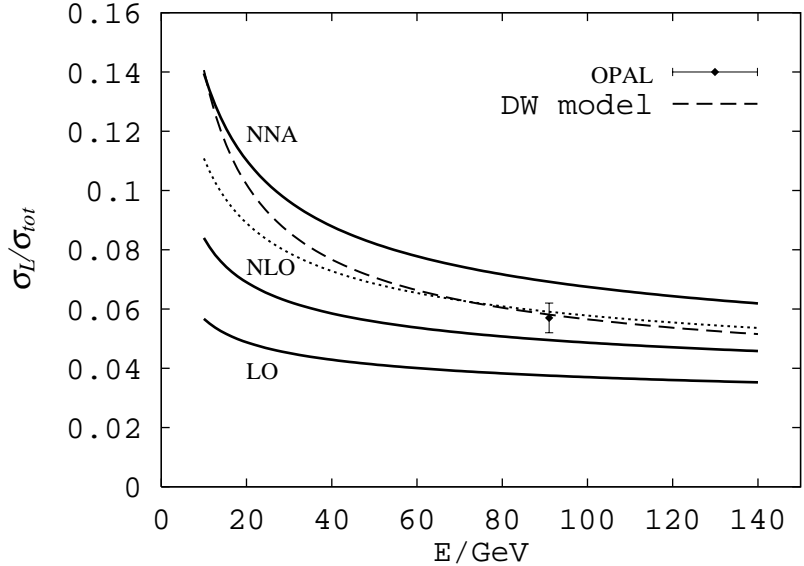}}
   \vspace*{0cm}
\caption{\label{figsigmaL}
Longitudinal  fraction in the total
$e^+e^-$ cross section calculated with
$\alpha_s(M_Z)=0.118$. 
Solid lines: leading order (LO), next-to-leading order (NLO) and
resummation of all orders in $\beta_0^n\alpha_s^{n+1}$ corrected for
the exact $O(\alpha_s^2)$ coefficient (NNA). 
Dashed line is the prediction of the Dokshitser-Webber model and dotted 
line is the average of NLO and NNA calculations.}
\end{figure}

Apart from measurements at a fixed energy, measuring the $Q$ dependence 
of $\sigma_L$ would be extremely interesting, especially at moderate 
energies. For other event shapes, it is known that a sizeable 
`hadronization correction' must be added to second-order perturbation 
theory in order to reproduce the energy dependence. On the other 
hand, once higher-order corrections are computed, for example in an 
approximation such as `naive non-abelianization', such hadronization 
corrections must be reconsidered, since the sum of higher-order corrections 
can already produce a steeper energy dependence. 

To illustrate this point, we have plotted
in Fig.~\ref{figsigmaL} 
the energy dependence of the total longitudinal cross section,
showing the leading-order (LO), next-to-leading order (NLO) and the 
resummed (NNA) results. We note the pronounced energy dependence 
of the resummed result towards lower energies. 
It is noteworthy that Monte Carlo parton showers do not produce 
such an energy dependence (see Fig.~5 of \cite{OPAL}) and resum 
a different set of higher-order corrections. At the same time, 
since higher-orders in $(\beta_0\alpha_s)^n$ are inseparable from 
$1/Q$ power corrections, it is plausible that a substantial part of the 
conventional hadronization correction effectively parametrizes these 
higher-order corrections.

A similar conclusion can be drawn from a comparison of the resummed 
NNA result with the prediction obtained in the approach of 
Dokshitser and Webber (DW) \cite{DOK95}, 
compare the dotted and the dashed curves in Fig.~\ref{figsigmaL}. 
The Dokshitser-Webber approach assumes universality of $1/Q$ power 
corrections and parametrizes them by an effective coupling. The 
dashed curve has been obtained following the prescription of 
\cite{DOK95} with the effective coupling fitted from the 
average $1-T$ ($T$ is thrust) and the relative coefficient of $1/Q$ 
for $1-T$ and $\sigma_L$ determined in the massive gluon scheme. In 
both cases the perturbative expansions are truncated at second order. 
It is seen that the energy dependence of the resummed perturbative 
results is not too different in NNA and the DW model.  
There is no conflict between the procedure of \cite{DOK95} and the 
resummation 
presented here, if the phenomenological $1/Q$ correction effectively 
parametrizes the higher-order perturbative contributions added 
in our approach.  If universality of power corrections holds, 
these perturbative corrections would also be universal, at least 
asymptotically in large orders. However, from the point of view presented 
here, the universality assumption is not required, since 
higher-order corrections are in principle calculable for each 
observable.

%%%%%%%%%%%%%%%%%%%%
% SECTION 6
%%%%%%%%%%%%%%%%%%%%

\section{Conclusions}

In this paper we investigated power-suppressed corrections to 
fragmentation processes. Drawing on ideas derived from large-order 
perturbation theory and applied before with some success to 
deep-inelastic scattering \cite{DMW,STE96,DAS962}, we have 
modelled the $x$ dependence of the leading $1/Q^2$ power corrections. 
The method we use is consistent with the QCD light-cone expansion 
of Ref.~\cite{BB91} and can be considered as a model of the 
$x$ dependence of certain twist-four correlation functions, based 
on the assumption of `ultraviolet dominance' of higher-twist 
matrix elements. The theoretical status of this assumption, as discussed 
in Sect.~2, is rather dubious --- eventually experimental results 
should clarify its relevance, as in the case of deep-inelastic 
scattering. 

For the first time we were able to numerically analyse additional 
ambiguities that affect the method for observables where the hadronic 
final state is weighted or not fully inclusive. These additional 
ambiguities affect power corrections to event shapes in general 
\cite{NAS95} and also shed some light on the reliability of 
`massive gluon calculations'. In Sect.~3.5 we provided a parametrization 
of the $x$ dependence of the leading ($1/Q^2$) power corrections, 
obtained by extracting the generic dependences on $x$ from the comparison 
between different implementations of the method. 
The parametrization depends on four constants which, although of order 
unity, should be determined experimentally. We should mention that 
we did not address kinematic effects due to the finite mass of the 
final-state hadron, which should be treated separately, as is done for 
target-mass effects in deep-inelastic scattering.

{}From the theoretical point of view, fragmentation processes look 
like a promising place to pursue further outstanding questions related 
to the existence and interpretation of $1/Q$ power corrections in processes 
without operator product expansion, such as Drell-Yan production or event 
shapes. We have seen that the light-cone expansion breaks down 
when $x\sim \Lambda/Q$ and quantities such as moments of the 
fragmentation cross section, which include the region of such small 
$x$, do not have a light-cone expansion. This situation is quite 
different from deep-inelastic scattering. In particular, the 
integrated longitudinal and transverse cross sections have a 
$1/Q$ correction. Fractional power corrections occur in fractional 
moments and in general the power behaviour of $x$-integrated 
quantities depends on the details of how the small-$x$ region 
is weighted. 

\bigskip
{\bf Acknowledgements.} We are grateful to Bryan Webber for discussions 
and for informing us on \cite{DAS96} prior to publication. 
M. B. thanks Stan Brodsky and Lance 
Dixon for interesting discussions and the Institute for Nuclear 
Theory in Seattle for its hospitality while part of this work was 
done. V.~B. is grateful to the Institute for Nuclear 
Theory in Seattle and the CERN Theory Group for their hospitality.  

%%%%%%%%%%%%%%%%%%%%%%%%%%%%%%%%%%%%%%%%%%%%%%%%%%

\end{document}